\documentclass[aip,pof,graphicx,12pt]{revtex4-1}
\usepackage{amssymb,amsbsy,times,fancyhdr,color}
\usepackage{amsmath}
\usepackage{mathrsfs,amsfonts}
\usepackage{latexsym,afterpage}
\usepackage{graphicx,subfigure,multirow}
\usepackage{color}
\newcommand{\dfd}{\mathrm{d}}
\newcommand{\imi}{\mathrm{i}}
\begin{document}




%\preprint{}

\title{On the necessary conditions for bursts of convection within the rapidly rotating cylindrical annulus} %Title of paper








%\author{Robert J. Teed, Chris A. Jones \& Rainer Hollerbach}
\author{Robert J. Teed}
\email[]{R.J.Teed@leeds.ac.uk}

\author{Chris A. Jones}

\author{Rainer Hollerbach}



%\thanks{}


\affiliation{Department of Applied Mathematics, University of Leeds, Leeds, LS2 9JT, UK}





%\date{\today}

\begin{abstract}

Zonal flows are often found in rotating convective systems. Not only are these jet-flows driven by the convection, they can also have a profound effect on the nature of the convection. In this work the cylindrical annulus geometry is exploited in order to perform nonlinear simulations seeking to produce strong zonal flows and multiple jets. The parameter regime is extended to Prandtl numbers that are not unity. Multiple jets are found to be spaced according to a Rhines scaling based on the zonal flow speed, not the convective velocity speed. Under certain conditions the nonlinear convection appears in quasi-periodic bursts. A mean field stability analysis is performed around a basic state containing both the zonal flow and the mean temperature gradient found from the nonlinear simulations. The convective growth rates are found to fluctuate with both of these mean quantities suggesting that both are necessary in order for the bursting phenomenon to occur.

\end{abstract}


\maketitle %\maketitle must follow title, authors, abstract and \pacs



\section{Introduction}
\label{sec:intro}

An interest in zonal flows originates from a desire to better explain various phenomena observed in geophysical and astrophysical bodies. These large azimuthal flows found in the atmospheres of the gas giants as well as planetary cores are thought to be driven by the interaction of convection and rotation. Jupiter, for example, has a clear banded structure of jets, made up of alternating prograde and retrograde zonal flows \cite{lim86,por03}. This pattern extends over the whole planet and the zonal flows are considerably stronger than the radial convection. Although the convection in both the deep Jovian atmosphere and the Earth's outer core will be affected by their respective magnetic fields, an understanding of the non-magnetic problem can provide insight to the physical structures observed. The depth to which the zonal flows extend in Jupiter's atmosphere is not known, though there is evidence to suggest that flows are considerably weaker in the core compared with the surface \cite{stajon02}. \citeauthor{bus76}\cite{bus76} suggested a model for convection in the Jovian atmosphere where zonal flows are not confined to the surface. The difficulties in modeling the interiors of the major planets has been discussed by \citeauthor{yan98}\cite{yan98}.

The linear theory of convection in spheres and spherical shells has now been comprehensively investigated. \citeauthor{rob68}\cite{rob68} and \citeauthor{bus70}\cite{bus70} derived some of the basic principles and the rapid rotation limit was discussed by \citeauthor{jonsow00}\cite{jonsow00} and \citeauthor{dor04}\cite{dor04}. However, performing three-dimensional nonlinear simulations in spherical geometry can be computationally expensive. Quasi-geostrophic models \citep{bus70,giljon06,rot07} 
assume that the rapid rotation leads to columnar structures with little
$z$-dependence, leading to two-dimensional models. The Busse annulus  \cite{bus70,busor86,bus86,orbus87,schbus92} is one such 
quasi-geostrophic model. It replicates several key aspects of convection 
in spherical geometry; for example, convection occurs in the form of tall thin columns which 
onset as thermal Rossby waves \citep{busor86}. Of particular relevance is the 
nonlinear model's ability to develop large zonal flows which may have a multiple jet structure \cite{jonabd03}.

Zonal flows have been found both in laboratory experiments \cite{buscar76,man96,aub01} 
and nonlinear simulations in the annulus \citep{bruhar93,jonabd03,rotjon06} and in the more physically realistic spherical shell geometry \cite{gil77,gil78a,gil78b,zha92,tilbus97,grobus01,chr01,chr02,bus02,hei05}.
Simulations of rotating convection in spherical shells were undertaken by \citeauthor{gil77}\cite{gil77,gil78a,gil78b} and \citeauthor{zha92}\cite{zha92} which produced zonal flows. More recent simulations \cite{tilbus97,aur01,chr01,chr02,bus02,hei05,hei07} have produced strong zonal flows, driven by the Reynolds stresses, with Rossby numbers of the correct order of magnitude. Interestingly, both steady and oscillatory solutions were found resulting in the discovery of a `bursting 
phenomenon' \cite{grobus01}. The bursting phenomenon, investigated within the annulus model by 
\citeauthor{rotjon06}\cite{rotjon06}, and in a quasi-geostrophic model taking
account of the curvature of the boundaries by \citeauthor{mordor04}\cite{mordor04} refers to the observation that convection can occur as short-lived bursts rather than the system evolving into a quasi-steady equilibrium. These bursts of convection are currently thought to be a result of a competition between the zonal flow and the convection. The convection
drives up the zonal flow strongly, but this zonal flow eventually disrupts the convection, which
then cannot sustain the zonal flow. The zonal flow dies away, allowing convection to occur
again, and repeat the cycle.

The three-dimensional simulations discussed above often do not produce a multiple jet structure of the zonal flow. The reason is that to get multiple jets very large rotation 
rates (very low Ekman numbers)
are required \cite{jonabd03}. Due to numerical difficulties the fully three-dimensional models have 
often been unable to achieve the rotation rate required, though in some exceptionally 
high-resolution three-dimensional simulations multiple jets have been found \cite{hei05,hei07}.

One of the attractions of the annulus model, as a simplified model for convection in the Jovian atmosphere, lies in its ability to produce both multiple jets and the bursts of convection. However, these properties are dependent on the boundary conditions. In general, stronger zonal flows and bursting are produced when stress-free top and bottom boundaries are imposed \cite{gil78b,bruhar93,chr01} whereas no-slip boundaries are able to generate a multiple jet 
structure \cite{jonabd03}. Similar results were found in the quasi-geostrophic model 
by \citeauthor{mordor06}\cite{mordor06}.
 The work of \citeauthor{rotjon06}\cite{rotjon06} shows that multiple jets and bursting appear to be mutually exclusive when the Prandtl number is unity.

The aims of this paper are three-fold. Firstly we wish to extend the simulations performed by \citeauthor{rotjon06}\cite{rotjon06} to parameter regimes where the Prandtl number is not unity. Secondly, we wish to check the consistency of our results with the Rhines scaling theory \cite{rhi75}. Thirdly, we wish to investigate what role the nonlinear interaction of temperature fluctuations have in the generation of bursts of convection since a mean temperature gradient is known to evolve when performing simulations. Note that the zonal flow in this work is generated by the nonlinear Reynolds stress, rather than by a thermal wind, for which the dynamics is rather different \cite{tee10}. To aid the reader we summarize in table \ref{tab:means} several quantities that appear in the article.

\begin{table}[tbhp]
\centering
\begin{tabular}{cccc}
\hline
Quantities & Section & Definition & Explanation \\
\hline\hline
$\bar{U}(y)$ & \ref{sec:nl} & $\langle u_x \rangle_x$ & $x$-averaged velocity (zonal flow) from nonlinear simulations \\
$\bar{\theta}(y)$ & \ref{sec:nl} & $\langle \theta \rangle_x$ & $x$-averaged (mean) temperature from nonlinear simulations \\
$\bar{\theta}^\prime(y)$ & \ref{sec:nl} & $\langle \theta^\prime \rangle_x$ & $x$-averaged (mean) temperature gradient from nonlinear simulations \\
$\bar{U}_{\text{max}}$ & \ref{sec:bursting} & $\max(\bar{U}(y))$ & Maximum value that $\bar{U}(y)$ takes in the domain \\
$\bar{U}_{\text{min}}$ & \ref{sec:bursting} & $\min(\bar{U}(y))$ & Minimum value that $\bar{U}(y)$ takes in the domain \\
$\bar{\theta}^\prime_{\text{max}}$ & \ref{sec:bursting} & $\max(\bar{\theta}^\prime(y))$ & Maximum value that $\bar{\theta}^\prime(y)$ takes in the domain \\
$\bar{\theta}^\prime_{\text{min}}$ & \ref{sec:bursting} & $\min(\bar{\theta}^\prime(y))$ & Minimum value that $\bar{\theta}^\prime(y)$ takes in the domain \\
$U_0(y)$ & \ref{sec:nlinlin} &  & Zonal flow included in the basic state of the linear theory \\
$G_0(y)$ & \ref{sec:nlinlin} &  & Mean temperature included in the basic state of the linear theory \\
\hline
% in \ref{sec:nllinzf} - \ref{sec:nllinzftg} 
\end{tabular}
\caption[]{Definition and explanation of quantities used in this article along with the section for which they first appear.}
\label{tab:means}
\end{table}

%%%%%%%%%%%%%%%%%%%%%%%%%%%%%%%%%%%%%%%%%%%%%%%%%%%%%%%%%%%%%%%%%%%%%%%%%%%%%%%%%%%%%%%%%%%%%%%%%%%%%%%%%%%%%%%%%%%%%%%
\section{Mathematical setup}
\label{sec:setup} 

We consider a fluid filled cylindrical annulus with inclined bounding surfaces for the top and bottom lids, see figure 1. The mean radius of the annulus is $r_0$, the gap between the two cylinders, referred to as the width, is $D$, and the height of the annulus at the outer cylindrical wall is $L$. The annulus rotates about the axial direction with angular velocity $\Omega$ and a temperature difference of $\Delta T$ is maintained between the two walls such that the outer and inner walls are at temperatures $T=0$ and $T=\Delta T$ respectively. We also take the gravity force to be acting radially inward and the annular end walls make an angle $\chi$ to the horizontal.

The natural choice of coordinate system for the annulus model would be cylindrical polar coordinates: $(r,\phi,z)$. However, by making the small-gap approximation of $D/r_0\ll 1$, the curvature terms of cylindrical polars can be neglected and Cartesian coordinates can be used. The azimuthal coordinate 
is $x$ and it increases eastwardly (acting like $\phi$) and $0\le y\le D$ is the radial coordinate (acting like $-r$). The axial coordinate, $z$, remains unchanged from cylindrical polars and ranges from $-L/2$ to $L/2$. Consequently, gravity acts in the positive $y$-direction and the direction of rotation is in the $z$-direction so that $\mathbf{g}=g\mathbf{\hat{y}}$ and $\mathbf{\Omega}=\Omega\mathbf{\hat{z}}$. 
The no penetration condition at the sloped end walls of the annulus is dependent on the inclination, $\chi$, so that
\begin{equation}
\cos(\chi)u_z\mp\sin(\chi)u_y\mp U_E=0\quad\textrm{on}\quad z\pm L/2.
\label{eq:endwallbc}
\end{equation}
Here $U_E$ is an Ekman suction or `bottom friction' term derived using the theory described by \citeauthor{gre68}\cite{gre68}. The purpose of the term is to replicate the effects of the Ekman boundary layer that arises when rigid boundaries are implemented. Hence $U_E$ is present \emph{only} when no-slip rather than stress-free boundaries are required.

\begin{figure}[ht]
\begin{center}
\includegraphics[height=6cm]{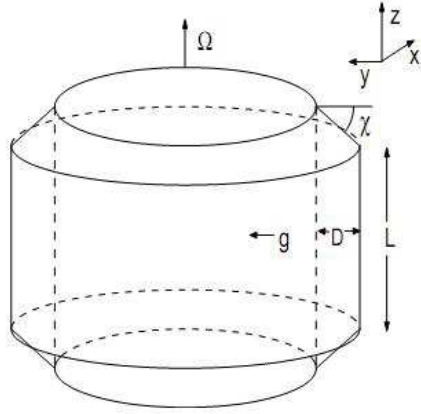}
\caption[Diagram depicting the physical setup of the Busse annulus]{Diagram depicting the physical setup of the Busse annulus; reproduced from \citeauthor{abd00}\cite{abd00}.}
\label{fig:annulus}
\end{center}
\end{figure}

The linear theory of the annulus model was originally discussed and solved by \citeauthor{bus70}\cite{bus70,busor86}. We use the $z$-component of the vorticity equation, which is
\begin{equation}
\frac{\partial \zeta}{\partial t} + \mathbf{u}\cdot\nabla \zeta - 2\Omega\mathbf{\hat{z}}\cdot\frac{\partial\mathbf{u}}{\partial z} = -g\alpha \frac{\partial T}{\partial x} + \nu\nabla^2 \zeta,
\label{eq:annvort1}
\end{equation}
where $\zeta$ is the $z$-component of the vorticity. Here we have neglected the 
$(\boldsymbol{\zeta}\cdot\nabla)\mathbf{u}$ term that usually appears in the vorticity equation since we are interested in the small Rossby number limit of rapid rotation where the planetary vorticity $2{\Omega}$ dominates over the fluid vorticity $\zeta$. This is the standard practice in the annulus model as well as other quasi-geostrophic models \cite{giljon06}. In the annulus model the term is of order $\chi$, which is taken to be much smaller than unity, while the other nonlinear terms are of order 1 as discussed by \citeauthor{busor86}\cite{busor86}.

We perturb around the basic state to acquire a similar set of equations to those of Busse, including nonlinear terms. 
We write  $T=T_0 + \theta$, where $T_0= y \Delta T /d$ is the conduction state profile,
and assume that $\chi\ll 1$. Hence the boundaries are nearly flat, the flow is nearly 
geostrophic and the $z$-component of the velocity is small compared with the horizontal components. This allows us to 
make the ansatz
\begin{equation}
\mathbf{u} = -\nabla\times\psi(x,y)\mathbf{\hat{z}} + u_z\mathbf{\hat{z}},
\end{equation}
where the vertical velocity, $u_z$, is a small ageostrophic part of the flow of order $\chi$. We substitute the perturbed forms of the the fields into equation (\ref{eq:annvort1}) as well as the relevant heat equation and integrate over $z$ applying the conditions of equation (\ref{eq:endwallbc}) at the sloped boundaries. We also non-dimensionalize using the length scale $D$, the viscous timescale $D^2/\nu$ and the temperature scale $\nu\Delta T/\kappa$. 
%We also suppose that our zonal flow has a typical velocity of $U^*$. 
This gives
\begin{gather}
\frac{\partial\nabla^2\psi}{\partial t} + \frac{\partial (\psi,\nabla^2\psi)}{\partial (x,y)} - \beta\frac{\partial\psi}{\partial x} = -Ra\frac{\partial\theta}{\partial x} - C|\beta|^{1/2}\nabla^2\psi + \nabla^4\psi, \label{eq:psieq2} \\
Pr\left(\frac{\partial\theta}{\partial t} + \frac{\partial(\psi,\theta)}{\partial(x,y)}\right) = -\frac{\partial\psi}{\partial x} + \nabla^2\theta,
\label{eq:thetaeq2}
\end{gather}
where the Jacobian, $\partial(A,B)/\partial(x,y)=(\partial_xA)(\partial_yB)-(\partial_xB)(\partial_yA)$ for some functions $A$ and $B$, has been introduced.
We have eliminated the vorticity by noting that $\zeta=\nabla^2\psi$. The beta parameter, $\beta$, Prandtl number, $Pr$, and Rayleigh number, $Ra$, are defined as
\begin{equation}
\beta=\frac{4\chi\Omega D^3}{\nu L},\qquad Pr=\frac{\nu}{\kappa},\qquad Ra=\frac{g\alpha\Delta TD^3}{\nu\kappa}.
\label{eq:annparas}
\end{equation}
In the annulus model the beta parameter effectively acts as an inverse Ekman number and therefore in the limit of rapid rotation we expect $\beta$ to be large. The small angle assumption has allowed us to write $U_E=-D^{3/2}\zeta(\chi/\beta L)^{1/2}$ for the Ekman suction \cite{jon07} and the term (in equation (\ref{eq:psieq2})) resulting from this bottom friction contains the parameter $C=(D/L\chi)^{1/2}$. Since the bottom friction is a phenomenon associated only with \emph{rigid} boundaries we explicitly set $C=0$ when implementing stress-free boundary conditions.

Since we have used the boundary conditions on the sloped end walls in order to integrate $z$ out of the problem, the only boundaries left to consider are those at the inner and outer walls of the cylinders. Equations (\ref{eq:psieq2} - \ref{eq:thetaeq2}) form a sixth order system of equations and thus we require six boundary conditions at $y=0$ and $y=1$. As well as there being no penetration we also demand these boundaries to be stress-free and have constant surface temperature so that
\begin{equation}
u_y=0\Rightarrow \frac{\partial\psi}{\partial x}=0, \qquad \frac{\partial u_x}{\partial y} = 0 \Rightarrow \frac{\partial^2\psi}{\partial y^2}=0, \qquad \theta = 0,
\label{eq:bcs}
\end{equation}
at $y=0$ and $y=1$.

%%%%%%%%%%%%%%%%%%%%%%%%%%%%%%%%%%%%%%%%%%%%%%%%%%%%%%%%%%%%%%%%%%%%%%%%%%%%%%%%%%%%%%%%%%%%%%%%%%%%%%%%%%%%%%%%%%%%
\section{Nonlinear results}
\label{sec:nl}

We perform nonlinear simulations of the equations (\ref{eq:psieq2}-\ref{eq:thetaeq2}). The system has four input parameters: $Pr,\beta,Ra$ and $C$. We integrate these nonlinear equations forward in time using a pseudo-spectral collocation method \cite{boy01}. We expand $\psi$ and $\theta$ using a Fourier and sine expansion in the $x$ and $y$-directions respectively. We therefore write
\begin{gather}
\psi(x,y,t) = \frac{1}{2}\sum_{l=-N_x}^{N_x}\sum_{m=1}^{N_y-1} \hat{\psi}_{lm}(t) e^{-\imi lx(2\pi/L_x)}\sin(m\pi y), 
\label{eq:psiexp1} \\
\theta(x,y,t) = \frac{1}{2}\sum_{l=-N_x}^{N_x}\sum_{m=1}^{N_y-1} \hat{\theta}_{lm}(t) e^{-\imi lx(2\pi/L_x)}\sin(m\pi y), \label{eq:thetaexp1}
\end{gather}
where we note that the choice of a sine expansion implicitly satisfies the boundary conditions of equation (\ref{eq:bcs}).
Since $\psi$ is real, we have that $\hat{\psi}_{-lm} = \hat{\psi}^*_{lm}$, where $*$ denotes complex conjugate.

The simulations are performed by implementing the Crank-Nicolson method to all but the nonlinear terms and a second-order Adams-Bashforth scheme to the nonlinear terms. Hence we use a semi-implicit method with only the nonlinear terms being calculated explicitly. 
We define `mean quantities' as follows. The zonal flow, $\bar{\mathbf{U}}$, is the $x$-average of the azimuthal component of the velocity, and $\bar{\theta}$ is the mean temperature so
\begin{equation}
\bar{\mathbf{U}} =  \bar{U}\mathbf{\hat{x}} = \langle u_x \rangle_x \mathbf{\hat{x}} = 
-\frac{\partial\langle\psi\rangle_x}{\partial y} \mathbf{\hat{x}}, 
\qquad\textrm{and}\qquad \bar{\theta} = \langle\theta\rangle_x.
\label{eq:Ubardef}
\end{equation}
 The $x$-average is defined as $\langle A\rangle_x = (1/L_x)\int_0^{L_x} A\dfd x$, for a scalar quantity, $A$. The only contribution to the mean quantities comes from modes with $l=0$ so that
\begin{equation}
\bar{U} = - \frac{1}{2}\sum_{m=1}^{N_y-1} m\pi\hat{\psi}_{0m}\cos(m\pi y) \qquad\textrm{and}\qquad \bar{\theta} = 
 \frac{1}{2}\sum_{m=1}^{N_y-1}\hat{\theta}_{0m}\sin(m\pi y).
\label{eq:zfexp}
\end{equation}

Zonal flow generation is governed by\cite{rotjon06} 
\begin{equation}
\frac{\partial\bar{U}}{\partial t} = -\frac{\partial}{\partial y}\langle u_xu_y\rangle_x -C|\beta|^{1/2}\bar{U} + \frac{\partial^2\bar{U}}{\partial y^2}.
\label{eq:zfgen}
\end{equation}
We note that zonal flow can be created by the Reynolds force, confirming that $\bar{U}$ is a nonlinear phenomenon, and destroyed by the friction terms. The addition of the bottom friction term is expected therefore to dampen the zonal flow; however, as discussed in section \ref{sec:intro}, we expect it to increase the likelihood of multiple jet solutions arising. Also of interest are the total kinetic energy and the zonal part of the kinetic energy, defined by
\begin{gather}
E_T = \frac{1}{L_x}\int(\nabla\psi)^2 \dfd S = \frac{1}{8}\sum_{l=0}^{N_x}\sum_{m=1}^{N_y-1}\left(\frac{4\pi^2 l^2}{L_x^2}+
m^2\pi^2\right)\vert\hat{\psi}_{lm}\vert^2, \qquad\textrm{and} \label{eq:Etotal1} \\
E_Z = \frac{1}{L_x}\int(\langle\nabla\psi\rangle_x)^2 \dfd S = \frac{1}{8}\sum_{m=1}^{N_y-1} m^2\pi^2\vert\hat{\psi}_{0m}\vert^2, \label{eq:Ezonal1}
\end{gather}
respectively.

In table \ref{tab:runs} we list the runs performed, which lie in the range $0.2\le Pr \le 5$ and $10^3 < \beta < 10^6$. 
$L_x$ is set at $2 \pi$, which is sufficiently large since the structures in
the $x$ direction have short wavelengths. In the previous work \cite{jonabd03,rotjon06} 
only Prandtl number unity was considered. We perform runs with the Rayleigh number 2.5, 2.75, 5 and 10 times that of the critical Rayleigh number for a given $Pr$ and $\beta$ as indicated in table \ref{tab:runs}. The rapid rotation approximation \cite{busor86} to the critical Rayleigh number for the Busse annulus 
\begin{equation}
Ra_c = \frac{3\beta^{4/3} Pr^{4/3}}{2^{2/3}(1+Pr)^{4/3}},
\label{eq:Racbusse}
\end{equation}
is adequate at these high $\beta$.

\begin{table}[tbhp]
\centering
\begin{tabular}{cccccccccc}
\hline
Run & $Pr$ & $\beta$ & $C$ & $Ra/Ra_c$ & $\tau$ & $m_*$ & Range of $\bar{U}_{\text{max}}$ 
& Range of $\bar{\theta}^\prime_{\text{max}}$ & Bursting\\
\hline\hline
%I & $1$ & $3.16\times 10^4$ & $0.316$ & $2.5$ & $14.5532$& No\\
%II & $1$ & $3.16\times 10^5$ & $0$ & $2.5$ & $0.7377$ & Yes\\
%III & $1$ & $3.16\times 10^5$ & $0.316$ & $2.5$ & $0.5509$ & Yes\\
I & $1$ & $7.07\times 10^3$ & $0.316$ & $2.5$ & $13.12$ & $2$ & $2-37$  & $1.2-3.2$ & No \\
II & $1$ & $7.07\times 10^5$ & $0$ & $2.5$ & $1.65$ & $2$ & ${587-679}$  & $0.9-1.9$ & Yes\\
III & $1$ & $7.07\times 10^4$ & $0$ & $2.5$ & $3.63$ & $1$ & ${319-534}$ & $0.2-3.2$ & Yes\\
IV & $1$ & $7.07\times 10^4$ & $0.00316$ & $2.5$ & $1.23$ & $2$ & ${210-298}$ & $0.5-2.5$ & Yes\\
%VII & $1$ & $7.07\times 10^4$ & $0.0316$ & $2.5$ & $3.3178$ & No\\
V & $1$ & $7.07\times 10^4$ & $0.316$ & $2.5$ & $8.08$ & $3$ & $12-156$ & $2.2-4.1$ & No \\
VI & $1$ & $7.07\times 10^5$ & $0.316$ & $2.5$ & $2.68$ & $5$ & $401-521$ & $2.2-3.3$ & No\\
\hline\hline
VII & $1$ & $5\times 10^5$ & $0$ & $2.75$ & $3.29$ & $2$ & ${834-1176}$ & $0.4-3.2$ & Yes \\
VIII & $1$ & $5\times 10^5$ & $0$ & $5$ & $0.93$ & $2$ & ${2591-4701}$ & $1.0-7.7$ & Yes \\
IX & $1$ & $5\times 10^5$ & $0.05$ & $2.75$ & $1.44$ & $3$ & $764-921$ & $0.85-2.9$ & No\\
X & $1$ & $5\times 10^5$ & $0.5$ & $2.75$ & $3.38$ & $5$ & $371-430$ & $3.4-5.0$ & No\\
XI & $0.5$ & $5\times 10^5$ & $0$ & $2.75$ & $2.69$ & $2$ & $903-1046$ & $0.35-1.05$ & No\\
XII & $0.5$ & $5\times 10^5$ & $0$ & $5$ & $0.25$ & $1$ & ${3993-5489}$ & $0.1-6.5$ & Yes\\
XIII & $0.5$ & $5\times 10^5$ & $0.5$ & $2.75$ & $2.64$ & $4$ & $372-570$ & $1.5-2.5$ & No\\
XIV & $2$ & $5\times 10^5$ & $0$ & $2.75$ & $2.11$ & $2$ & ${646-873}$ & $0.9-4.6$ & No\\
XV & $2$ & $5\times 10^5$ & $0$ & $5$ & $1.03$ & $2$ & ${1422-2387}$ & $1.3-12.6$ & Yes\\
XVI & $2$ & $5\times 10^5$ & $0.05$ & $2.75$ & $2.42$ & $3$ & $580-678$ & $2.8-4.4$ & Yes\\
XVII & $2$ & $5\times 10^5$ & $0.5$ & $2.75$ & $2.47$ & $7$ & $40-134$ & $6.5-8.1$ & No\\
XVIII & $5$ & $5\times 10^5$ & $0$ & $2.75$ & $4.69$ & $2$ & $321-385$ & $6.0-8.5$ & No\\
XIX & $5$ & $5\times 10^5$ & $0$ & $5$ & $0.56$ & $2$ & $1094-1167$ & $8.0-12.1$ & No\\
XX  & $0.2$ &  $5\times 10^5$ & $0$ &  $2.75$ & $2.01$ & $1$ & $897-1089$ & $0.1-0.7$ & No\\
XXI & $0.2$ & $5\times 10^5$ & $0.05$ & $2.75$ & $2.75$ & $4$ & $162-241$ & $0.3-0.6$ & No\\
XXII & $0.2$ & $5\times 10^5$ & $0.5$ & $2.75$ & $2.20$ & $6$ & $201-294$ & $0.5-0.8$ & No\\
XXIII & $0.2$ & $5\times 10^5$ & $0$ & $5$ & $0.43$ & $1$ & $3004-3389$ & $0.1-1.1$ & Yes\\
XXIV & $1$ & $5\times 10^5$ & $0$ & $10$ & $0.56$ & $1$ & $8467-11784$ & $0.1-19.6$ & Yes\\
XXV & $0.5$ & $5\times 10^5$ & $0$ & $10$ & $0.29$ & $1$ & $947-1203$ & $0.1-16.9$ & Yes\\
\hline
\end{tabular}
\caption[Table displaying the parameter sets used for the various nonlinear runs]{Table displaying the parameter sets
 used for the various nonlinear runs. Also indicated are: the total integration time, $\tau$, the dominant 
wavenumber, $m_*$ and the ranges of the maxima of the zonal flow and mean temperature gradient, and whether 
bursting is seen or not.}
\label{tab:runs}
\end{table}

Each of the runs displayed in table \ref{tab:runs} is integrated until a quasi-steady or quasi-periodic state has evolved from the initial condition. A random initial state is used for each run. For the parameter values considered, we find that the final
state is independent of the initial conditions. The quantity $\tau$, appearing in table \ref{tab:runs}, represents the total number of time units that the particular run was integrated over. Also in table \ref{tab:runs} we display $m_*$, which denotes the time-averaged dominant radial wavenumber. The value of $m_*$ determines whether multiple jets are present; a solution has $m_*+1$ jets and we define $m_*\geq 3$ to denote a multiple jet solution. In table \ref{tab:runs} we also indicate the range of the maxima of $\bar{U}$ and $\bar{\theta}^\prime$ in order to show typical flow strength and temperature gradients for each run. 
Runs where bursting occurs are noted, bursting being defined as solutions having quasi-periodic
time-dependence. We have predominantly used the resolution $(N_x,N_y)=(256,128)$ although runs VIII, XIV, XVI and XVII have $(N_x,N_y)=(384,128)$ and runs XV, XVIII, XIX, XXIII, XXIV and XXV have $(N_x,N_y)=(512,128)$.

\subsection{Previous work}

We briefly review previous work, runs I-VI being for parameters regimes examined by \citeauthor{jonabd03}\cite{jonabd03} and \citeauthor{rotjon06}\cite{rotjon06}. As $\beta$ is increased, disturbances become smaller in the $x$-direction in line with 
the scaling $k\sim\beta^{1/3}$ predicted by the linear theory \cite{busor86}. However, linear theory predicts thin
disturbances with a simple $\sin \pi y$
dependence in the $y$-direction, whereas nonlinear effects make the dominant wavenumber in $y$ similar to that
in $x$, see figure \ref{fig:contours}. There is also an increase in the strength of the zonal flow as $\beta$ is increased; compare the magnitude of $\bar{U}$ in table \ref{tab:runs} for runs III and II where $C=0$ or alternatively for runs V and VI, where $C\ne0$. %Since the forcing is the same for all of these runs ($Ra=2.5Ra_c$), there must be another explanation for the differing magnitudes of the zonal flow. 
Recall from equation (\ref{eq:zfgen}) that the magnitude of the zonal flow is determined by the balance of the Reynolds forcing against the frictional terms. At larger $\beta$ the streamlines slope more and give rise to an increased Reynolds forcing 
and larger zonal flow even though the magnitudes of $u_x$ and $u_y$ in equation (\ref{eq:Ubardef}) are not
much increased. The general increase in the magnitude of the zonal flow must saturate at some large value of $\beta$ since the sloping of the streamlines cannot continue indefinitely.

The introduction of the bottom friction has two main consequences. Firstly, the zonal flow is weakened as expected from equation (\ref{eq:zfgen}). For runs III and V, which have the same value of $\beta$ but different values of $C$ the flow is much weaker in the case where $C\ne 0$ (see table \ref{tab:runs}). The zonal flow has depleted in strength from $\approx400$, in run III, to $\approx70$ in run V. % Note also that there is far less order in the contour plots for $\psi$ and $\theta$ in figure \ref{fig:runV} since the zonal flow is weak. This is also the case in figure \ref{fig:runI}.
Secondly, the introduction of the Ekman layer drastically improves the likelihood of multiple jet solutions. The only runs, of these first six, where multiple jets are presented are runs V and VI. These two runs both have $C=0.316$, which is the largest value of $C$ tested, for these initial runs. For runs where $C=0$ we also do not find any evidence of multiple jets since runs II and III are dominated by wavenumbers $m=2$ and $m=1$ respectively (see table \ref{tab:runs}).  The possibility of multiple jets arising also increases as $\beta$ is increased. Thus, relatively large values of $C$ and $\beta$ are preferred for multiple jets resulting in run VI having the most jets (six in total) of any of these first six runs. The number of jets found for each run can be compared directly with those of table 1 from \citeauthor{jonabd03}\cite{jonabd03} and table 1 from \citeauthor{rotjon06}\cite{rotjon06}, where we see excellent agreement.

\subsection{Runs VII to XXV}
\label{sec:newruns}

\begin{figure}[p]
\vspace{-5mm}
%\begin{center}
%\hspace{-25pt}
\abovecaptionskip=-6pt
\addtolength{\subfigcapskip}{-20pt}
\addtolength{\subfigcapmargin}{-13pt}
\subfigure[tight][\ Run X: $Pr=1$, $C=0.5$, $Ra/Ra_c=2.75$.]{\label{fig:runX}\includegraphics[scale=0.48]{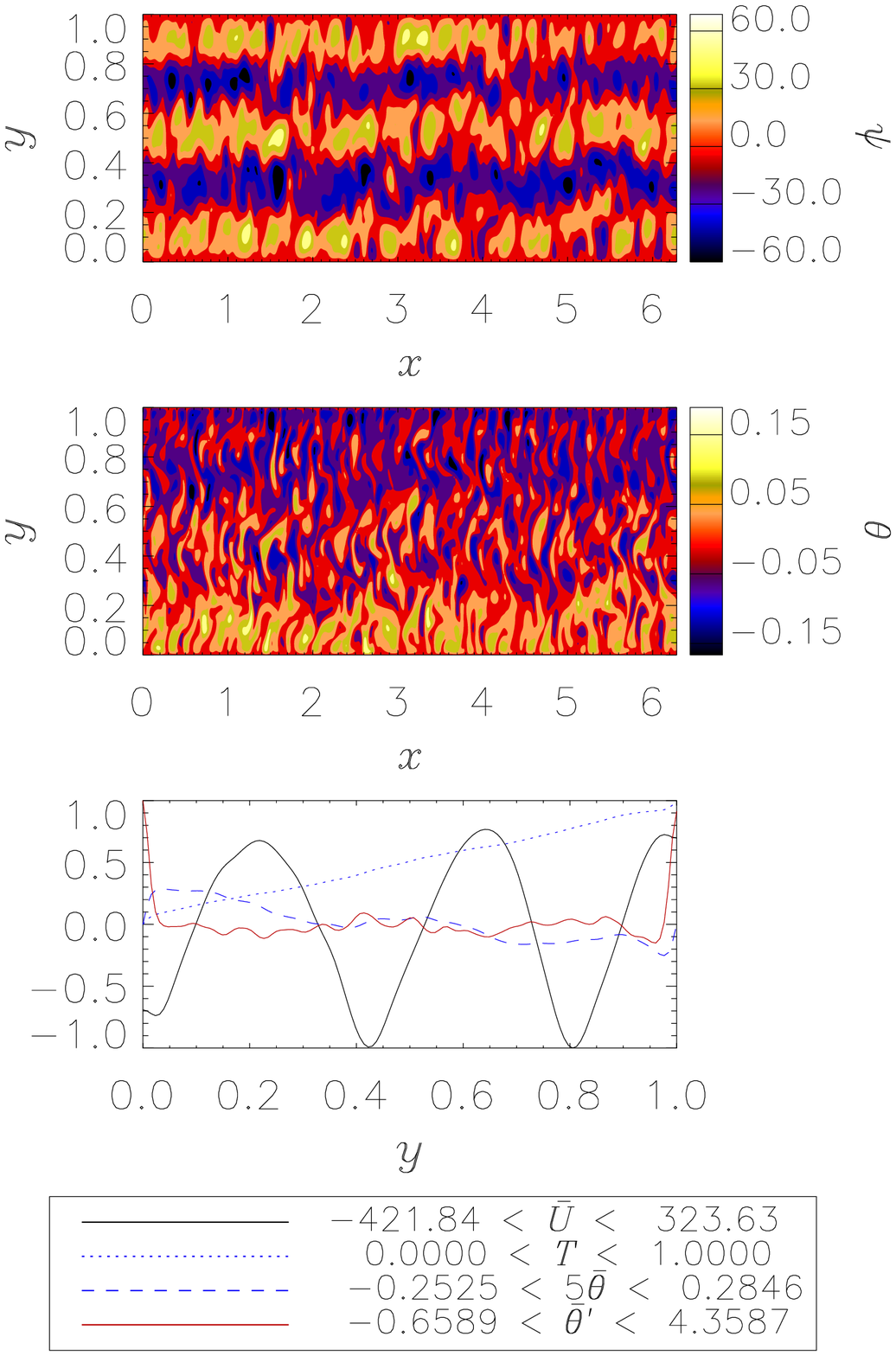}}
\hspace{0pt}
\subfigure[\ Run VII: $Pr=1$, $C=0$, $Ra/Ra_c=2.75$.]{\label{fig:runVII}\includegraphics[scale=0.48]{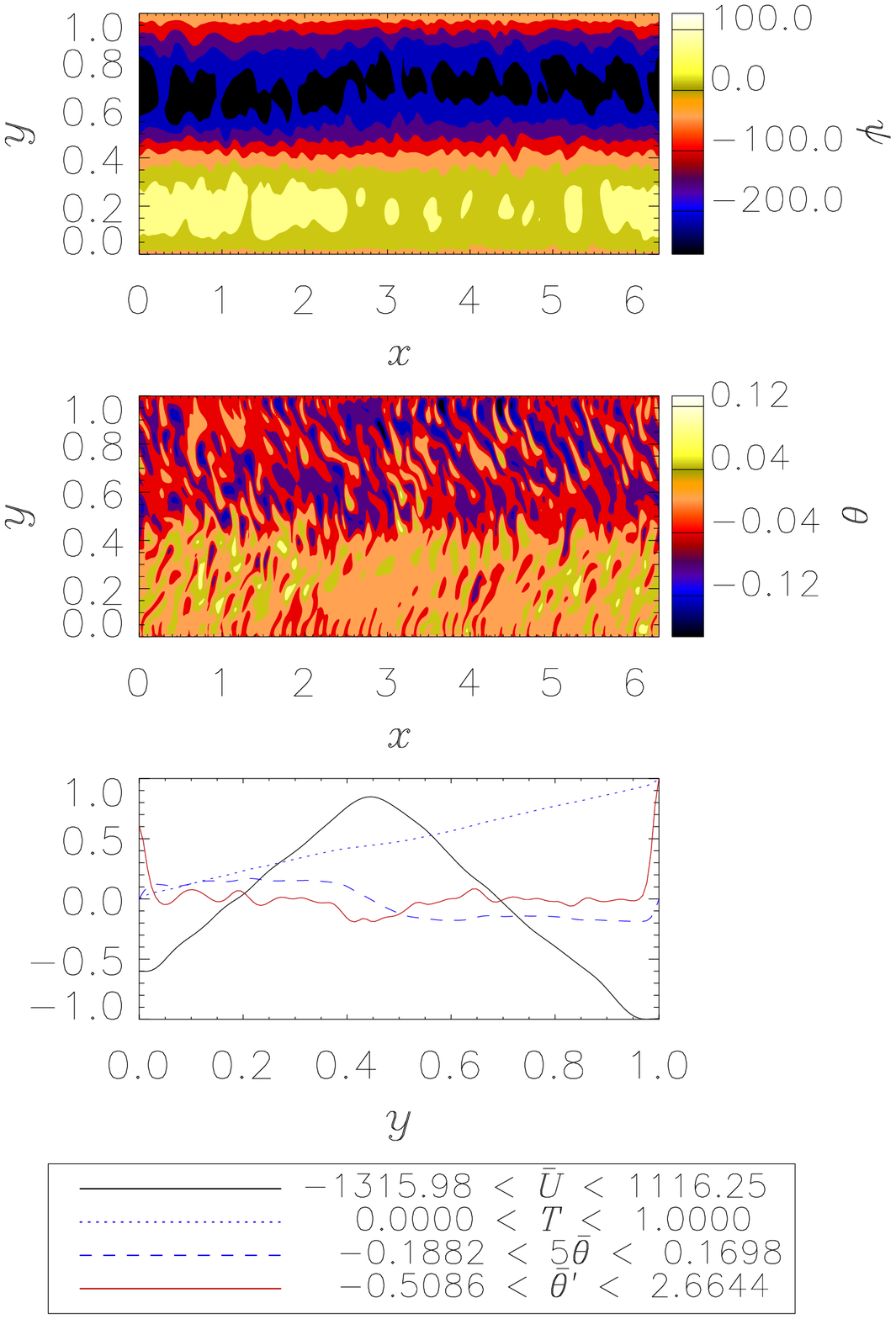}}
%\vspace{10pt}
\subfigure[\ Run XV: $Pr=2$, $C=0$, $Ra/Ra_c=5$.]{\label{fig:runXV}\includegraphics[scale=0.48]{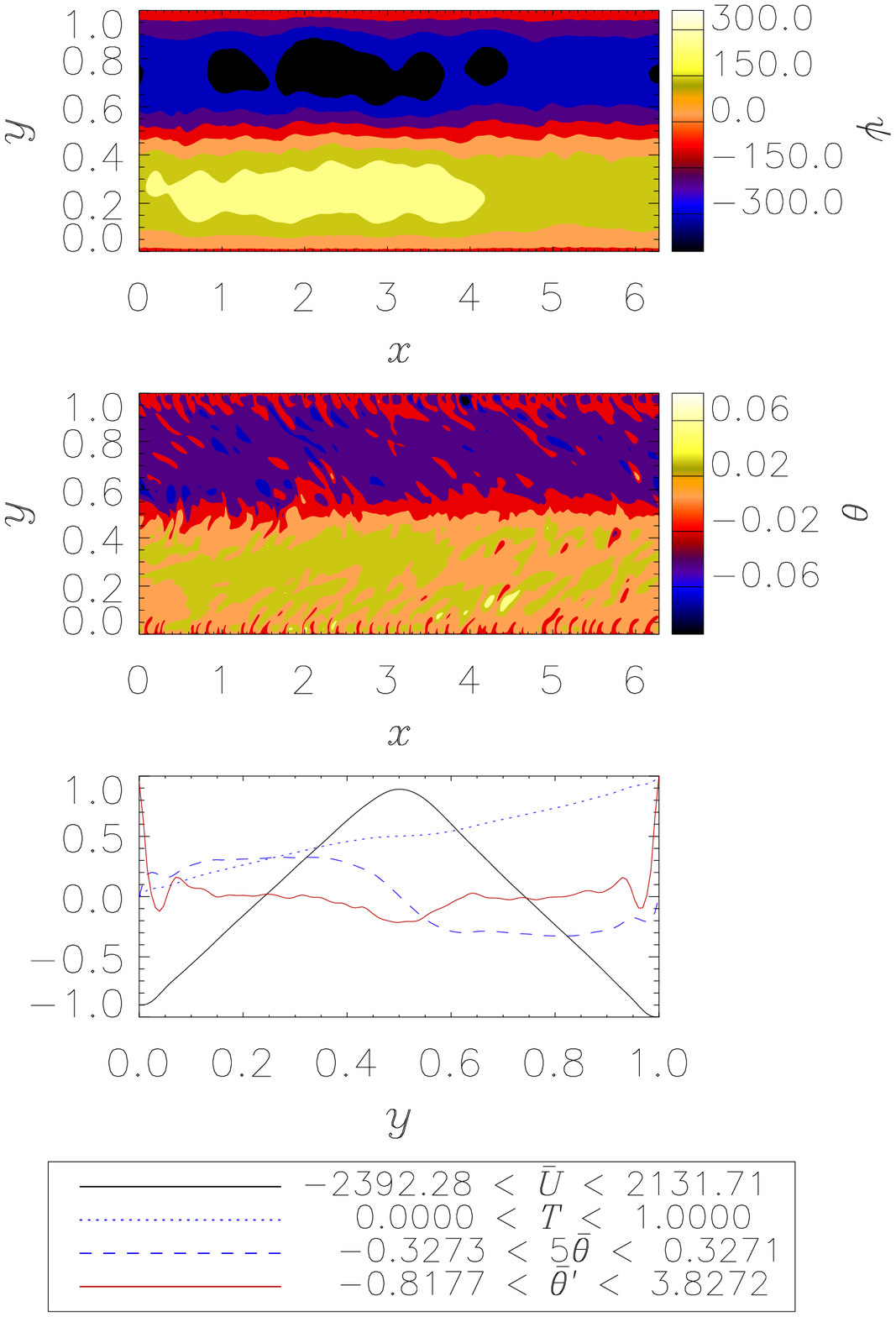}}
\hspace{0pt}
\subfigure[\ Run XVII: $Pr=2$, $C=0.5$, $Ra/Ra_c=2.75$.]{\label{fig:runXVII}\includegraphics[scale=0.48]{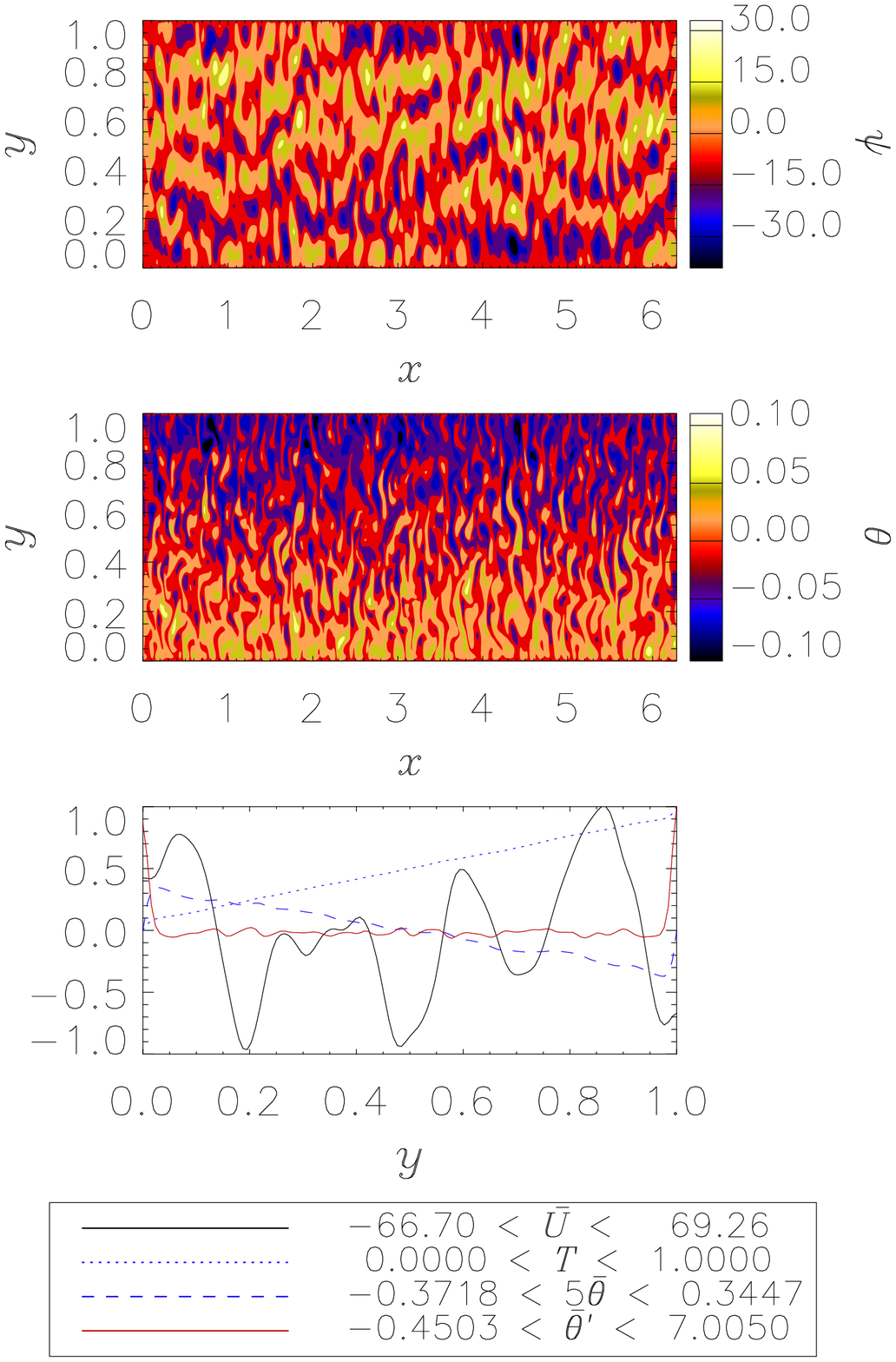}}
\vspace{10pt}
\caption[Contour plots for several runs. All plots have $\beta=5\times 10^5$.]{Contour plots for various runs. All plots have $\beta=5\times 10^5$.}
\label{fig:contours}
%\end{center}
\end{figure}

We explore the parameter space further, and the results are shown in runs VII to XXV. The parameter regimes used for these runs can, again, be found in table \ref{tab:runs}, where we see that all have $\beta=5\times10^5$. We have considered further values of the Prandtl number and Rayleigh number, whilst continuing to vary $C$. In figure \ref{fig:contours} we plot the state of the simulation for four particular runs that display differing behavior. Each plot is displayed at $t=\tau$, once a final state has been achieved. We should note that the plots for figures \ref{fig:runVII} and \ref{fig:runXV} are only \emph{snapshots} at a particular time since the final state of these solutions is time dependent. Conversely, the plots of figures \ref{fig:runX} and \ref{fig:runXVII} are typical of the final state since the solution is quasi-steady for runs X and XVII.

Three plots are displayed in each subfigure of figure \ref{fig:contours}. The top two plots display the $\psi$-contours and the $\theta$-contours at time $\tau$, respectively. In the case of the $\psi$-contours, positive and negative values represent clockwise and counter-clockwise motion respectively. In the third plot of each figure we plot four quantities: the zonal flow, $\bar{U}$, the mean temperature profile, $\bar{\theta}$, the \emph{total} temperature profile, $T=T_0+\bar{\theta}$, and the mean temperature gradient, $\bar{\theta}^\prime$. The values of $\bar{U}$ have been normalized using $\max(|\bar{U}|)$ and likewise, $\bar{\theta}^\prime$ has been normalized using $\max(|\bar{\theta}^\prime|)$. Also, the exact value of $T$ has been plotted, whereas $\bar{\theta}$ has been amplified by a factor of five in order to be more clearly displayed. The range over which the quantities vary are presented beneath the third plot.

%The first six runs are for parameter regimes used by \cite{jonabd03,rotjon06} so that $Pr=1$ and $Ra/Ra_c=2.5$ throughout.
%We are able to make some general observations about the dynamics that remain common among all runs performed. From 
%figure \ref{fig:contours} it is clear that the nonlinear effects have significantly altered the simple structure of the 
%fields predicted by the linear theory \cite{busor86} where $\psi$ and $\theta$ take the form of thin disturbances 
%stretching across the whole annular layer (that is, from $y=0$ to $y=1$). Although this structure is observed in 
%certain regions for some runs, the overall flow pattern is rather different to that predicted in the linear theory.

For runs with $m_*=2$ a net eastward zonal flow is produced at $y=1/2$ (see, for example, figure \ref{fig:runVII}). This is caused by the interaction of the predominantly clockwise motions for $y<1/2$ with the predominantly counter-clockwise motions for $y>1/2$. The resultant negative $y$-gradient in $\psi$ produces an eastward zonal flow ($\bar{U}>0$) as expected from equation (\ref{eq:Ubardef}). In some plots, for example the $\psi$-plot of figure \ref{fig:runXV}, the zonal flow is strong enough to dominate the dynamics so much that convective cell patterns are no longer visible. In such cases, the correlation between regions of strong zonal flow and regions of strong $\partial\psi/\partial y$ is very clear.

%There are also general observations that can be made from the $\theta$-plots of figure \ref{fig:contours}. 
%The $\sin(\pi y)$ dependence of the linear theory, where one would expect alternating yellow and blue vertical 
%stripes, has again been suppressed by nonlinear effects. In fact, we find that the preference is almost 
%exclusively for yellow 
%($\theta>0$) and blue ($\theta<0$) in the regions $y<1/2$ and $y>1/2$ respectively, which is a result of 
%the mean temperature, $\bar{\theta}$, attempting to `flatten out' the static temperature profile. 

%Figure 2

Many of the runs display a striking correlation of the $\theta$-contours with the slope of $\bar{U}$. The 
$\theta$-contours show the local slope of the flow because temperature is advected with the flow. 
This slope then gives the sign of the Reynolds stress, which via equation (\ref{eq:zfgen}) determines the form of the zonal flow.
Run X, displayed in figure \ref{fig:runX}, shows a multiple jet solution; in this case five jets are apparent located at the edges of the bands displayed in the $\psi$-plot. The $\theta$-contours show a \lq herring-bone\rq  pattern, as the slope of the convection alternates
in direction as the zonal flow alternates in sign. 

The final states achieved by runs VII and XV are much the same as evidenced by the similarity of the first and third plots of figures \ref{fig:runVII} and \ref{fig:runXV}. The difference in the $\theta$-plot is a result of the unsteady nature of these solutions. Both runs have settled into a bursting solution, however the snapshots of figures \ref{fig:runVII} and \ref{fig:runXV} are taken at different phases in a bursting cycle. We shall discuss this further in section \ref{sec:bursting}.

The pattern of the fields are also be affected by the Prandtl number. The convection rolls are larger at small Prandtl 
number and decrease in size at larger Prandtl number, which is consistent with the preferred wavenumber at onset
in the limit of rapid rotation \cite{busor86}: $k_c\sim (Pr/(1+Pr))^{1/3}$. It is interesting that even in this strongly
nonlinear, time-dependent convection, the convection rolls follow the predicted wavenumber of linear theory. The Prandtl 
number still plays an important role in the pattern of convection found.

A larger Prandtl number is also beneficial to multiple jet production. Runs X, XIII and XVII each have the same (large) value of $C$ but table \ref{tab:runs} shows that between four and seven jets are produced depending on the value of $Pr$. However, as we see from figure \ref{fig:runXVII}, the appearance of the seven jets in run XVII when $Pr=2$ is associated with a weak zonal flow. This results in the $\psi$-contours lacking a clear banded structure unlike in the equivalent case for $Pr=1$ (see figure \ref{fig:runX}). Thus it seems that increasing the Prandtl number causes the system to lose its banded structure at a lower value of $C$.

The mean zonal flow is larger at small Prandtl number, and weakens at large Prandtl number. At large Prandtl number the Reynolds
number of the convective flow is reduced, and this leads to a smaller zonal flow, see equation (\ref{eq:zfgen}). The
mean temperature gradients behave in the opposite manner, being weaker at low Prandtl number, because thermal
diffusion is then relatively more important than advection, see equation (\ref{eq:thetaeq2}). 
This dependence is evidenced in table \ref{tab:runs}; the most clear example is by comparing the ranges of $\bar{U}_{\text{max}}$ and $\bar{\theta}^\prime_{\text{max}}$ for runs VIII, XII and XV.

\subsection{Rhines scaling theory}

We now briefly consider the implications of the Rhines scaling theory \cite{rhi75} on our results. By suggesting that the predominant balance is between the inertial and Coriolis force terms, \citeauthor{rhi75} found that the length scale of the flow should scale as $(U^*/\beta)^{1/2}$ for a typical flow strength of $U^*$, and some evidence for this scaling in the context of the convective annulus
has been found \cite{jonabd03}. The value for $U^*$ that should be used has been a topic of considerable debate due to the existence of two main typical flow strengths: the convective velocity and the zonal flow strength. \citeauthor{rhi75} originally envisaged the turbulent eddy velocity, corresponding here to the convective velocity would be used and some models of Jupiter's jets still use this approach \cite{schliu09}. However an alternative view is that the zonal flow strength should be used, rather than the eddy velocity, and this has experimental \cite{gil07} and numerical support \cite{jonabd03,hei07}.

Here we consider the applicability of the scaling theory to our results for each type of velocity separately. The convective velocity, $U_C$, acts in the radial direction and thus we set $U_C=(\max\{u_y\}-\min\{u_y\})/2$ as a measure of the convective flow strength. Similarly we set the zonal velocity, $U_Z$, as $U_Z=(\max\{\bar{U}\}-\min\{\bar{U}\})/2$. In each case the relevant quantity is also averaged over an appropriate period of time. The scaling theory stipulates that the number of jets, $M$, satisfies
\begin{equation}
M = c\left(\frac{\beta}{U^*}\right)^{1/2},
\label{eq:rhines}
\end{equation}
for some constant scaling factor, $c$. Hence for each run VII to XXV (where $\beta=5\times10^5$) we are able to calculate the number of jets predicted by inserting either $U_C$ or $U_Z$ for $U^*$. The value of $c$ is chosen in order to best fit the actual results for the number of jets found from the numerical simulations; that is, the value of $m_*$.

\begin{figure}
\abovecaptionskip=-0pt
\addtolength{\subfigcapskip}{-10pt}
\subfigure[\footnotesize{\ Zonal velocity with $c=0.1345$.}]{\label{fig:rhineszon}
\includegraphics[scale=0.45]{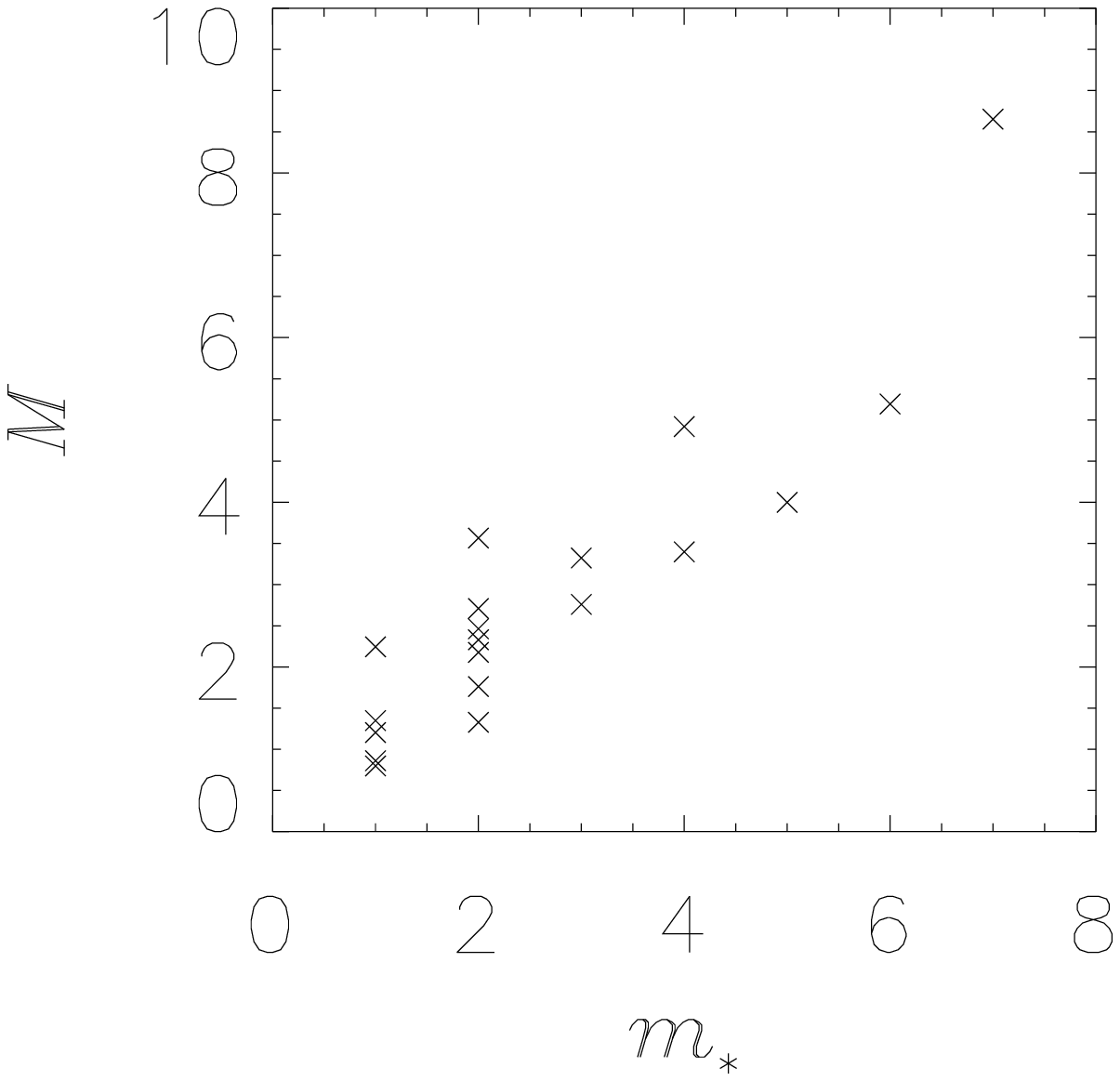}}
\hspace{30pt}
\subfigure[\footnotesize{\ Convective velocity with $c=0.1529$.}]{\label{fig:rhinescon}\includegraphics[scale=0.45]{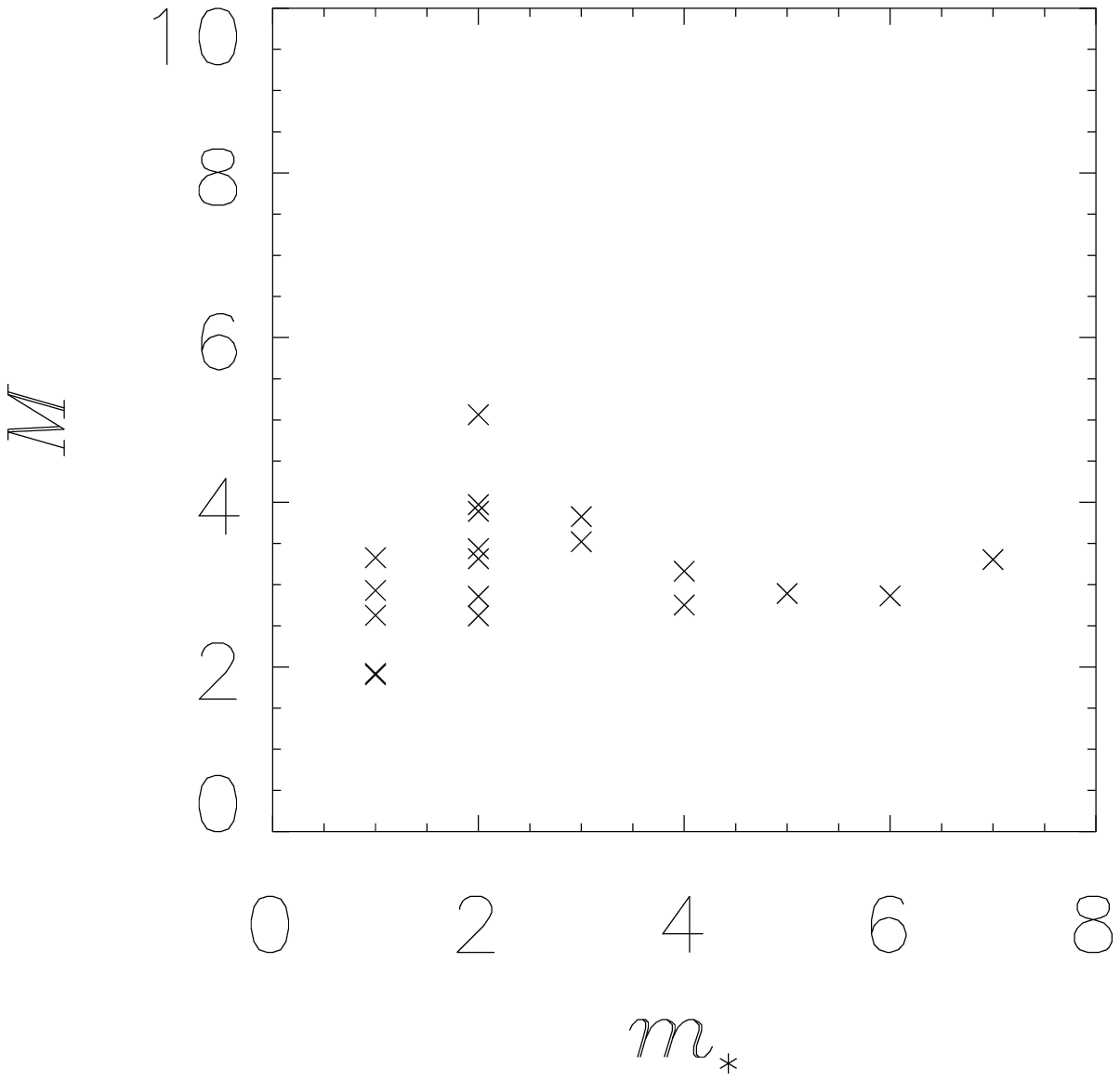}}
\caption{Plots of the true number of jets found from the simulations, $m_*$ against the predicted number of jets from the Rhines
 scaling theory, $M$, for runs VII to XXV.}
\label{fig:rhines}
\end{figure}

In figure \ref{fig:rhines} we plot the number of jets predicted using the Rhines scaling theory against the actual number of jets for the two cases: $U^*=U_Z$ and $U^*=U_C$. The value of the scaling factor used is given in the caption of each plot. Exact agreement between the theory and the simulation results would result in a line of best fit with $M=m_*$. Figure \ref{fig:rhineszon} shows a reasonably good fit indicating that the scaling theory may well be predicting the correct length scale when the zonal velocity is used. However, it is clear from figure \ref{fig:rhinescon} that the same is not true when the convective velocity is entered as the typical flow strength. In particular, the theory is unable to predict the number of jets accurately for simulations with a large number of jets. In fact, even at low $m_*$ the agreement is not as consistent as figure \ref{fig:rhineszon}. Consequently, we find that zonal velocities must be used in the Rhines scaling theory in order to best replicate the number of jets observed. However, due to the limited range of $\beta$ that we have tested, this conclusion really must be tested for simulations with larger rotation rates. We hope to perform this in future work.

%Figure 3
%\begin{figure}
%\includegraphics[scale=0.45]{rhineszon.ps}} \hspace{30pt}

When $\beta$ is held constant, equation (\ref{eq:rhines}) indicates that the typical flow strength must reduce in order to acquire a larger number of jets. As previously noted, multiple jets are associated with solutions that have weak zonal flow and hence the correct dependence on the flow strength is possible when $U^*=U_Z$. Conversely, the convective velocity remains at a near-constant strength regardless of the number of jets. Thus the line of best fit of figure \ref{fig:rhinescon} satisfies $M\approx$ constant and poor agreement is found between the predicted and actual number of jets.

We should note that \citeauthor{jonkuz09}\cite{jonkuz09} and \citeauthor{chr02}\cite{chr02}
found that implementing no-slip boundary conditions had the effect of removing multiple jets that were present under stress-free conditions. This is the opposite effect to that observed here, although a key difference is that their domain was spherical. \citeauthor{jonkuz09}\cite{jonkuz09} and \citeauthor{chr02}\cite{chr02} actually found that no-slip boundary conditions reduced zonal flow to the extent that it was indistinguishable from convective velocities, so that not only were multiple jets removed but so too was \emph{any} large scale zonal flow. Zonal flow production is more efficient in the annulus model compared with a spherical model. In the annulus model we have found that zonal flow is so strong with stress-free boundaries that the Rhines length fills the whole domain, precluding multiple jets. In the stress-free spherical shell models the zonal flow is relatively weaker, so according to the Rhines scaling theory \cite{rhi75} the Rhines length is smaller allowing multiple jets to form. Bottom friction in the annulus model reduces
the zonal flow, and hence the Rhines length, so that multiple jets can fit into the domain. No-slip boundaries in the spherical shell
models weaken the zonal flow so much that it is impossible to distinguish it from the chaotic convection. 
It is possible that multiple jets may reappear in the no-slip spherical shell models if the Ekman number is reduced
so that there is less bottom friction, but sufficiently small Ekman numbers are currently out of reach computationally.  
Similarly, multiple jets may appear in the stress-free annulus model at very large $\beta$ and moderate $Ra$, as the
Rhines length scales as $(U/ \beta)^{1/2}$. We have however been unable, so far, to reach the values of $\beta$ that may be required to observe this.

The Rhines scaling only applies when the convection is fully developed. At lower Rayleigh numbers, the bottom friction may allow multiple jets to appear because the Ekman friction term introduced when $C\ne0$ is a scale-independent damping term. Therefore, unlike the interior viscous diffusion which dampens the small-scales more greatly than large-scale structures, the Ekman friction `hits' all scales equally. This increases the likelihood of small-scale structures, such as multiple jets, appearing rather than just one large-scale equatorial jet. At Rayleigh numbers below twice critical with no bottom friction, multiple jets are damped out
by the interior viscosity, even though the zonal flow is small enough that equation (\ref{eq:rhines}) would predict
multiple jets. 

\subsection{The bursting phenomenon}
\label{sec:bursting}

\begin{figure}[p]
%\begin{center}
%\hspace{-25pt}
\addtolength{\subfigcapmargin}{-4pt}
\subfigure[\ Run XII: $Pr=0.5$, $C=0$, $Ra/Ra_c=5$.]{\label{fig:XIIe}\includegraphics[scale=0.38]{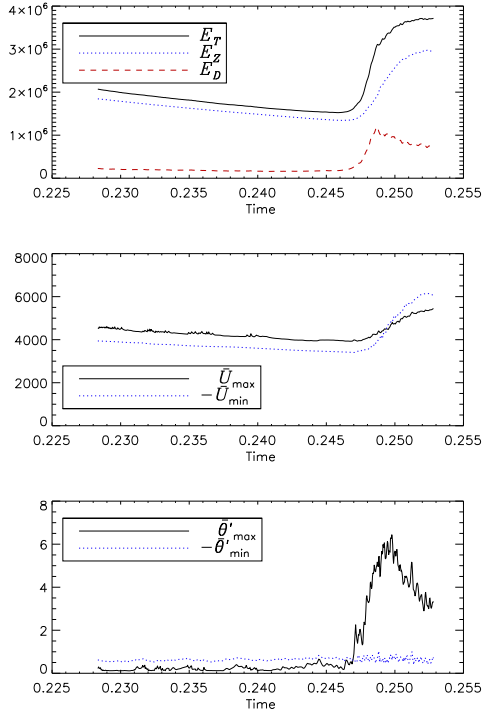}}
\hspace{20pt}
\subfigure[\ Run VII: $Pr=1$, $C=0$, $Ra/Ra_c=2.75$.]{\label{fig:VIIe}\includegraphics[scale=0.38]{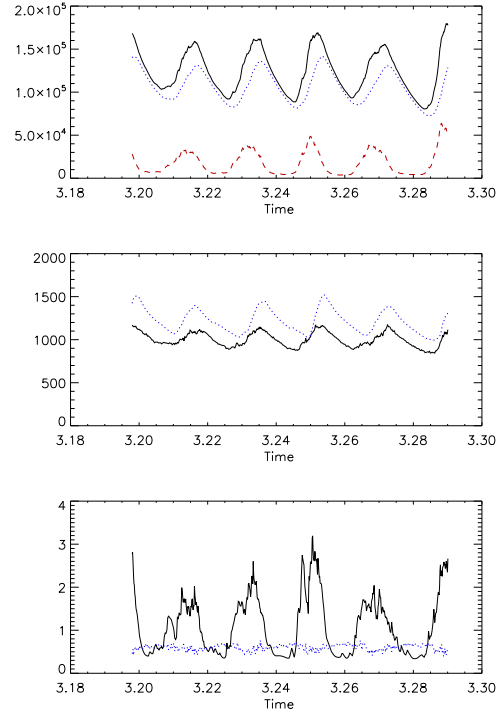}}
\subfigure[\ Run XV: $Pr=2$, $C=0$, $Ra/Ra_c=5$.]{\label{fig:XVe}\includegraphics[scale=0.38]{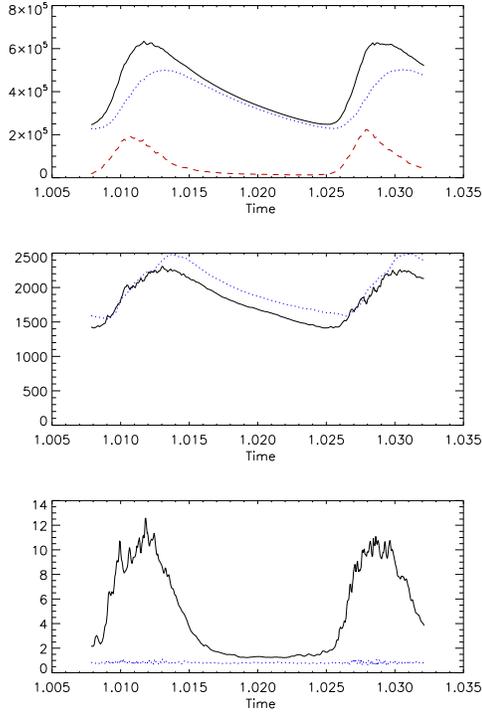}}
\hspace{20pt}
\subfigure[\ Run XVII: $Pr=2$, $C=0.5$, $Ra/Ra_c=2.75$.]{\label{fig:XVIIe}\includegraphics[scale=0.38]{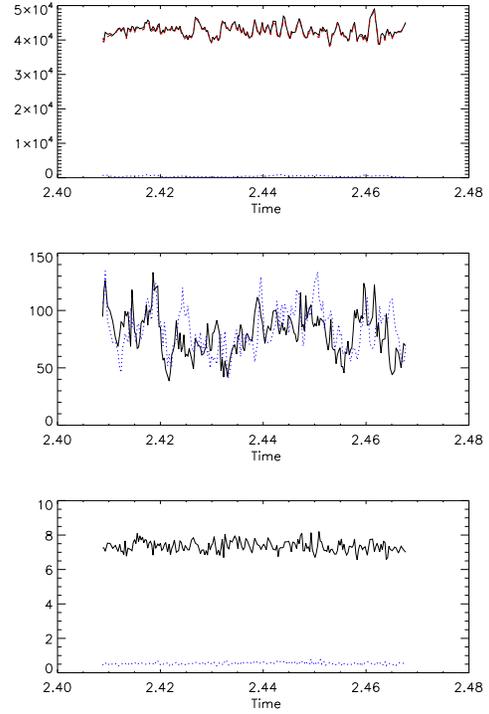}}
%\vspace{50pt}
\caption[Time evolution of the energy and mean quantity extrema plots for various runs]{Time evolution
of the energy and mean quantity extrema plots for various runs. The quantities plotted are defined in panel (a).}
\label{fig:energies}
%\end{center}
\end{figure}

For runs XII, VII, XV and XVII we plot, in figure \ref{fig:energies}, several more quantities as they evolve. In each 
figure \ref{fig:XIIe} to \ref{fig:XVIIe} the top plot displays the various energies: the total kinetic energy, $E_T$, the zonal kinetic energy, $E_Z$, and the difference between the two, $E_D$ (effectively the convective energy), which were defined by 
equations (\ref{eq:Etotal1} - \ref{eq:Ezonal1}). The remaining two plots contain the extremum values (that is, the maxima and minima) of the mean quantities, at each timestep. Figures \ref{fig:XIIe} to \ref{fig:XVIIe} allow us to observe the bursting phenomenon.
%Figure 4

Figure \ref{fig:VIIe}, which is for run VII, perhaps best showcases the bursts of convection, with several bursts apparent. A clear quasi-periodic phenomenon is occurring with all quantities displaying an oscillatory nature. The zonal flow is oscillating over a range of approximately 500. At times when there is a sharp increase in the energy and the extrema of $\bar{U}$, the zonal flow is driven up by the convection. However, the strong shear of the zonal flow then inhibits the convection, which depletes the source of energy for the zonal flow. Note that the maxima of the zonal energy occurs shortly after the maximum values of the extrema of $\bar{U}$. The zonal energy then decreases to a level that allows the convection to build up and a new burst can occur.

The physical mechanism by which the zonal flow actually suppresses the convection has not been much discussed. We have performed a linear stability analysis for the annulus model with a {\bf linear} flow pattern imposed in the basic state in a similar manor to that of \citeauthor{tee10}\cite{tee10}. We find that the critical Rayleigh number always increases with increasing flow strength confirming that zonal flow inhibits convection. We also find that the critical wavenumber is substantially reduced, so longer waves are preferred. 
The temperature perturbation of short wavelength cells is disrupted by the shear. In mathematical terms, the
large zonal flow means that the temperature perturbation $\theta$ in equation (\ref{eq:thetaeq2}) must be small for
the term $\bar{U} \partial \theta / \partial x$ to balance
the advection down the mean temperature gradient $- \partial \psi / \partial x$ and the temperature diffusion $\nabla^2 \theta$, but small temperature perturbations require very large $Ra$ to provide sufficient buoyancy. In practice, the system responds by choosing a longer wavelength parallel to the zonal flow (small $k$) to reduce the effect of the $\bar{U} \partial \theta / \partial x$ term, but this is not optimal for rapidly rotating convection which prefers large $k$. So the critical Rayleigh number is increased in the presence of shear. Note that this argument only applies to modes which are buoyancy driven, as modes which are driven by the shear itself do not rely on the temperature perturbation. However, modes driven by shear flow instability do not seem to play a big role in our simulations. Note also that it is shear which disrupts the temperature perturbation. If there is a large constant zonal flow $\bar{U}$, then waves with phase velocity $\approx \bar{U}$ can happily grow, but if the large velocity $\bar{U}$ varies with position, it is not possible for a single phase speed $c$ to cancel out $\bar{U}$ everywhere in equation (\ref{eq:thetaeq2}).

%We now move on to figures \ref{fig:runIXe} and \ref{fig:runXe} which are also for $Pr=1$ but bursting is less evident. In figure \ref{fig:runIXe}, for run IX with $C=0.05$,
Table \ref{tab:runs} indicates the runs for which bursting was observed with the range of the zonal flow also displayed. We find that as $C$ is increased from zero, in runs IX and X, bursting ceases and the range of the oscillations of the zonal flow is smaller (compare with run VII). Therefore, we can conclude that the bottom friction hinders the bursting phenomenon, which is in agreement with the previous work \citep{jonabd03,rotjon06}. For the runs where bursting occurs for $Pr=1$ (that is, runs VII, VIII and XXIV) the period of the bursting is found to be $\approx 0.02$ of a diffusion time. This can be observed from figure \ref{fig:VIIe}.

When $Pr=0.5$ we see, from table \ref{tab:runs} that the strength and range of the zonal flow is small for run XI where $Ra=2.75Ra_c$. However, when the Rayleigh number is increased to five times critical in figure \ref{fig:XIIe}, for run XII, the zonal energy forms the majority of the kinetic energy in the system. There is also evidence of the bursting phenomenon with a gradual decline in all of the quantities in the three plots before a sharp increase at $t\approx 2.48$. Bursting also continues to be found in run XXV where $Ra/Ra_c=10$. Therefore the possibility of convective bursts exists at $Pr=0.5$ so long as the driving is large enough.

In figures \ref{fig:XVe} and \ref{fig:XVIIe} we plot the energy and mean quantity extrema plots for runs XV and XVII where $Pr=2$. Figure \ref{fig:XVe} once again shows clear evidence of bursting, this time at five times critical, with significant fluctuations in both mean quantities. The maximum values of $E_T$ and $\bar{\theta}^\prime_\textrm{max}$ occur shortly before the peaks in $\bar{U}_{\textrm{max}}$ and $-\bar{U}_{\textrm{min}}$. The period of time between bursts has also remained constant at $\approx 0.02$ despite the increase in the Prandtl and Rayleigh numbers compared with run VII. This suggests that the period of the bursts may not be strongly dependent on either $Pr$ or $Ra$. From figure \ref{fig:XVe} it is clear that the snapshot for this run (see figure \ref{fig:runXV}) is taken during a time of strong zonal flow; that is a post-convective burst. The convection in figure \ref{fig:runXV} is also localized due to the strong zonal flow. This is in contrast to figure \ref{fig:runVII} which is taken \emph{during} a burst. This shows that during a bursting cycle there are both periods where convection occurs everywhere and where convection is localized. This is a common attribute of all bursting runs. Also of note is that the range of the fluctuation in the maximum value of the mean temperature gradient is larger than in the cases of lower Prandtl number (compare with figure \ref{fig:XIIe}). %We do not give a plot for run XVI; it is similar to figure \ref{fig:runIXe}. 
Figure \ref{fig:XVIIe}, for run XVII where $C=0.5$, again shows that increasing the bottom friction causes the bursting to halt, as well as reducing the magnitude of the zonal flow itself.

At $Pr=5$ we find that, despite the zonal energy forming the majority of the kinetic energy, the bursting ceases. The extremely small range of the zonal flow for run XVIII in table \ref{tab:runs} indicates that the values of these quantities are nearly constant over a long period of time. The 
same situation was found for run XIX, which has a larger Rayleigh number so bursting does not occur even for values of $Ra$ that are several times critical. No bursting was observed for runs with $Ra/Ra_c=2.75$ and $Pr=0.2$. With the non-zero values of $C$ used in runs XXI and XXII, the zonal flow is weak, so bursting would not be expected. In run XX, $C=0$ and the zonal flow is quite strong, but no bursting was found. However, increasing the Rayleigh number to five times critical in run XXIII, produces bursting although the oscillations are significantly weaker than those found for equivalent parameters in the $Pr=0.5$ case. This suggests that the onset of bursting is delayed as the Prandtl number is decreased.

In the annulus problem, bursting can be thought of as temporal intermittency, but in other geometries spatial intermittency can also occur. Spatial intermittency is sometimes referred to as `nests of convection' \cite{bro08}. Spatial intermittency  can occur in two different ways. It can occur with temporal intermittency, that is when the burst occurs it onsets preferentially in the neighborhood of a particular longitude. We see this happening in the annulus model, but typically the burst soon spreads out throughout the whole domain. More localized bursts have been seen in spherical shell geometry \cite{hei12,bal07}. However, in spherical shell geometry persistent nests of convection can occur, both in Boussinesq \cite{grobus01} and in anelastic convection \cite{bro08}. In this configuration, convection only occurs in patches which drift azimuthally in longitude, while individual convection columns drift through the patch, growing as they enter the patch and decaying as they leave it. We have not seen this phenomenon in the annulus model. In spherical geometry, the Rossby waves propagate faster in the outer parts of the shell, where the boundary slope is steeper, and more slowly in the deep interior where the slope is shallower. In the annulus model the boundary slope is constant, so this differential propagation speed with radius does not occur, which maybe why we did not find persistent nests of convection in the annulus.

%%%%%%%%%%%%%%%%%%%%%%%%%%%%%%%%%%%%%%%%%%%%%%%%%%%%%%%%%%%%%%%%%%%%%%%%%%%%%%%%%%%%%%%%%%%%%%%%%%%%%%%%%%%%%%%%%%%%%%%%%%%%%%%
\section{Mean field stability theory}
\label{sec:nlinlin}

In the previous section, we saw how large zonal flows and mean temperature gradients readily appeared 
under many parameter regimes. It is desirable to explore the disruptive effects that these mean 
quantities have on the convection in order to better explain how the bursts occur. Perhaps the most 
informative method is to consider a linear theory with the mean quantities derived from the nonlinear 
code used to
define a basic state from which the growth rates of convection can be observed. This is what we analyze 
in this section.

We consider a zonal flow, $\mathbf{u_0}=U_0(y)\mathbf{\hat{x}}$ and a mean temperature profile, $T_0 + G_0(y)$, 
to be included in the basic state. $G_0(y)$ measures the departure of the mean temperature from the
conduction state. Perturbations $\tilde \psi$ and $\tilde \theta$ around this basic state satisfy
\begin{gather}
\frac{\partial\nabla^2 {\tilde \psi}}{\partial t} + ReU_0\frac{\partial\nabla^2 {\tilde \psi}}{\partial x} 
- (\beta+ReU_0^{\prime\prime})\frac{\partial{\tilde \psi}}{\partial x} = -Ra\frac{\partial{\tilde \theta}}{\partial x} 
- C|\beta|^{1/2}\nabla^2{\tilde \psi} + \nabla^4{\tilde \psi}, \label{eq:psieqlin} \\
Pr\left(\frac{\partial{\tilde \theta}}{\partial t} + ReU_0\frac{\partial{\tilde \theta}}{\partial x} + 
\frac{\partial{\tilde \psi}}{\partial x}\frac{\dfd G_0}{\dfd y}\right) = 
-\frac{\partial{\tilde \psi}}{\partial x} 
+ \nabla^2{\tilde \theta},
\label{eq:thetaeqlin}
\end{gather}
with the introduction of terms involving $U_0$ and $G_0$. Disturbances have $\exp (ikx)$ dependence,
$k$ being the wavenumber in the $x$-direction. We have non-dimensionalized the zonal flow by assuming that it has a typical velocity, $U^*$, to give a Reynolds number
\begin{equation}
Re=\frac{DU^*}{\nu}.
\end{equation}

This linear problem is solved using the same method as in \citeauthor{tee10}\cite{tee10}. %Rather than having a \emph{static} basic state (that is, $\mathbf{u_0}=\mathbf{0}$ and $T_0=\Delta T y/D$) we have now introduced the effects of nonlinear terms into the basic state for both the velocity and temperature profiles. We do this in order to ascertain what effect the mean quantities have on the linear convective growth rates.
No assumption has yet been made regarding the form of $U_0$ or $G_0$. However, now we use the runs 
discussed in section \ref{sec:nl} to provide the mean quantities to be entered into the 
linear theory. Of course, as the system is evolved during these runs the zonal flow and 
mean temperature change at each timestep. In order to fully analyze the 
effects of the mean quantities on the linear theory we perform the linear stability 
analysis \emph{at each timestep}, 
which allows us to see how the growth rates of the linear system vary as the dynamics 
of the nonlinear system evolve. Therefore we add a subroutine to the nonlinear code, 
which solves the linear stability problem at each timestep. With the same parameter set 
as that being used in the nonlinear run and with $U_0(y)$ and $G_0(y)$ set 
equal to $\bar{U}$ and $\bar{\theta}$ respectively, the subroutine outputs the largest growth rate, 
as well as the corresponding frequency, $\omega$, and wavenumber, $k$.  We set $Re=1$ so that the magnitude of the zonal flow comes solely from the nonlinear simulations. We expect that the growth rate will be large when a burst arises. Conversely, when the zonal 
flow is strong the expectation is that the growth rate will attain a minimum due to the disruption of convection
by zonal flow as we discussed in section \ref{sec:bursting}. In order for 
convection to cease we expect to find \emph{marginal} growth rates at times of 
large $\bar{U}$. The idea we explore is that it is the small scale convection which drives 
the zonal flow and the mean temperature gradient, but for much
of the time these mean quantities are such that convective instability is suppressed. When 
these mean quantities have weakened through diffusive effects, we expect to see positive 
growth rates for convective instability, and the onset of a burst of convection. 

In the plots that we shall discuss, the growth rate, wavenumber and frequency will be functions 
of time. We are primarily interested in the growth rate of the fastest growing mode and how it 
varies as the nonlinear system is evolved. This is because we wish to ascertain if the magnitude 
of growth is at all correlated with the mean quantities. Consequently, we primarily look at the 
linear outputs for runs of section \ref{sec:nl} where the bursting phenomenon was witnessed. 
In particular, we discuss results from the linear theory for the same time intervals and runs as 
those taken for the plots in figure \ref{fig:energies}, in order to ease comparison.

\subsection{Linear results with nonlinear zonal flow}
\label{sec:nllinzf}

\begin{figure}[p]
%\begin{center}
%\hspace{-25pt}
\subfigure[\ Run XII.]{\label{fig:XIIzf}\includegraphics[scale=0.38]{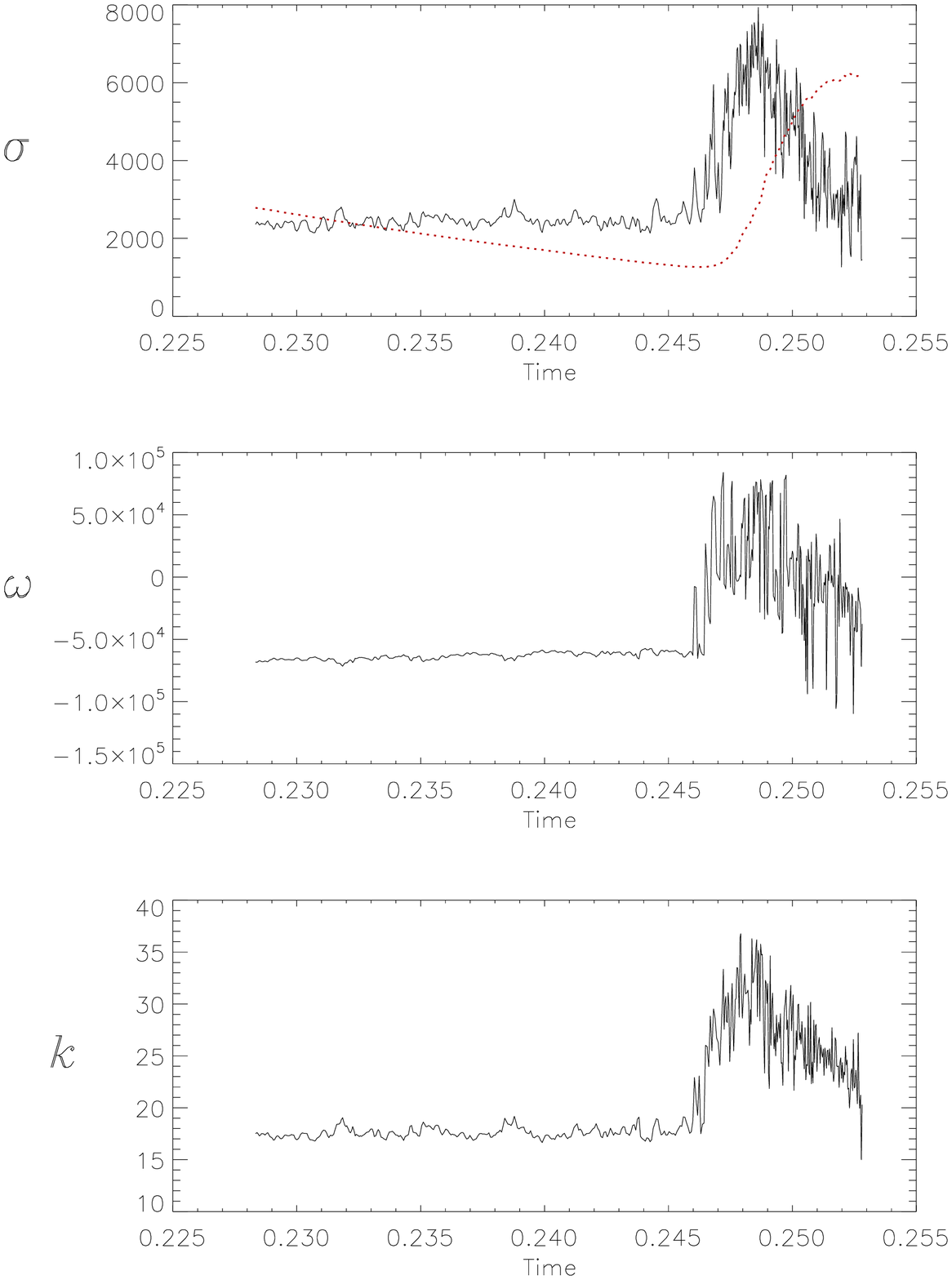}}
\hspace{30pt}
\subfigure[\ Run VII.]{\label{fig:VIIzf}\includegraphics[scale=0.38]{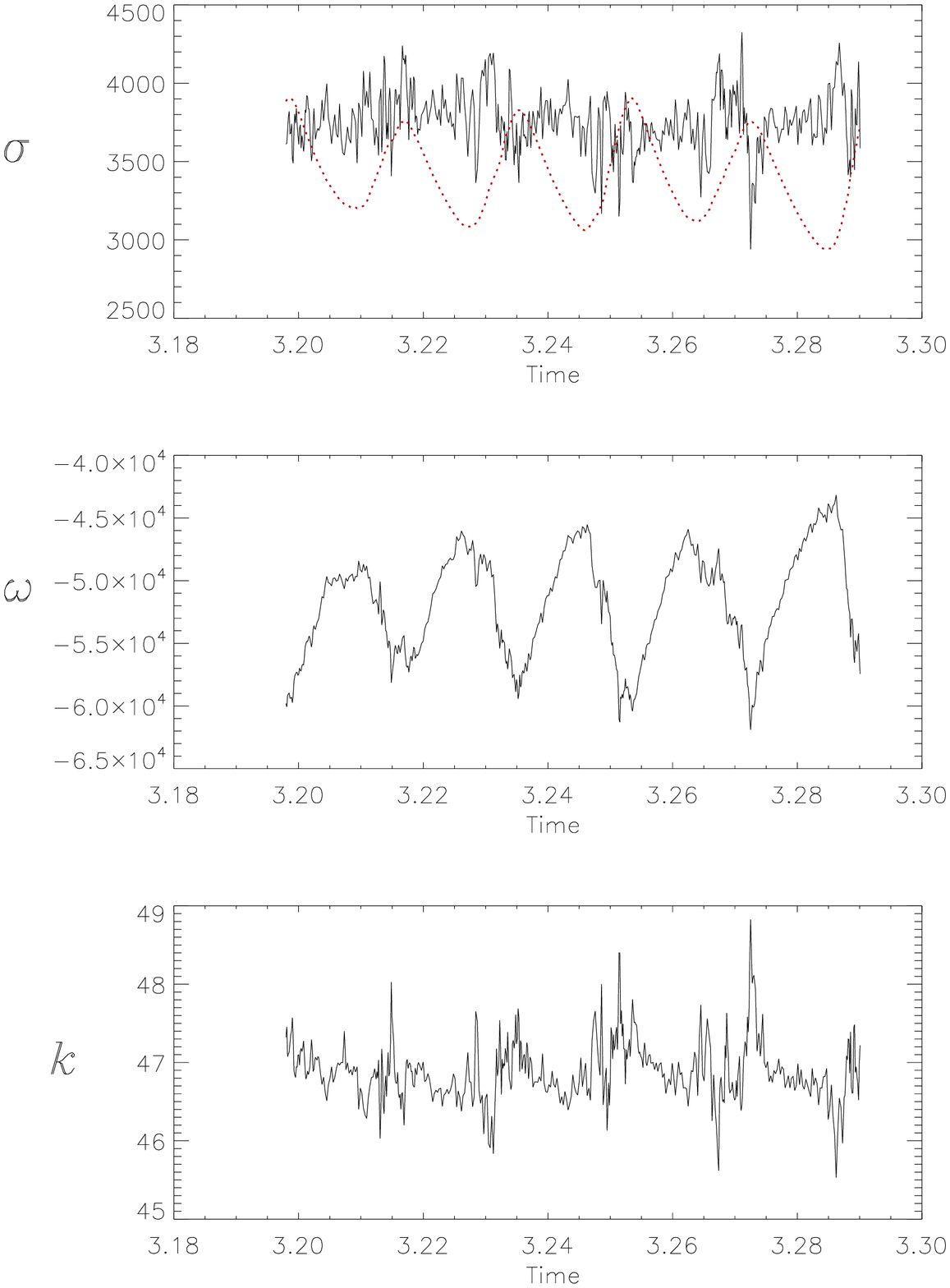}}
\subfigure[\ Run XV.]{\label{fig:XVzf}\includegraphics[scale=0.38]{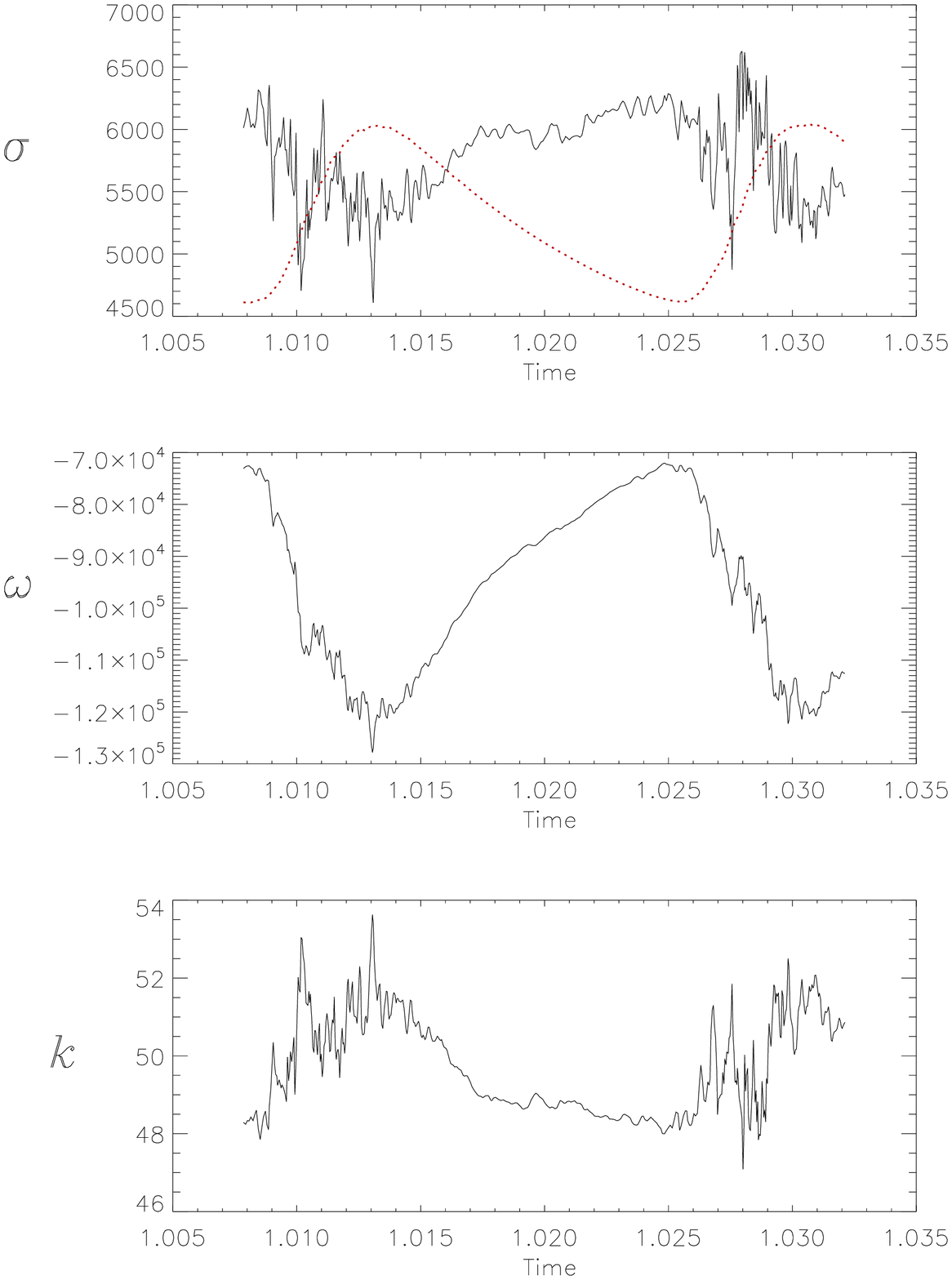}}
\hspace{30pt}
\subfigure[\ Run XVII.]{\label{fig:XVIIzf}\includegraphics[scale=0.38]{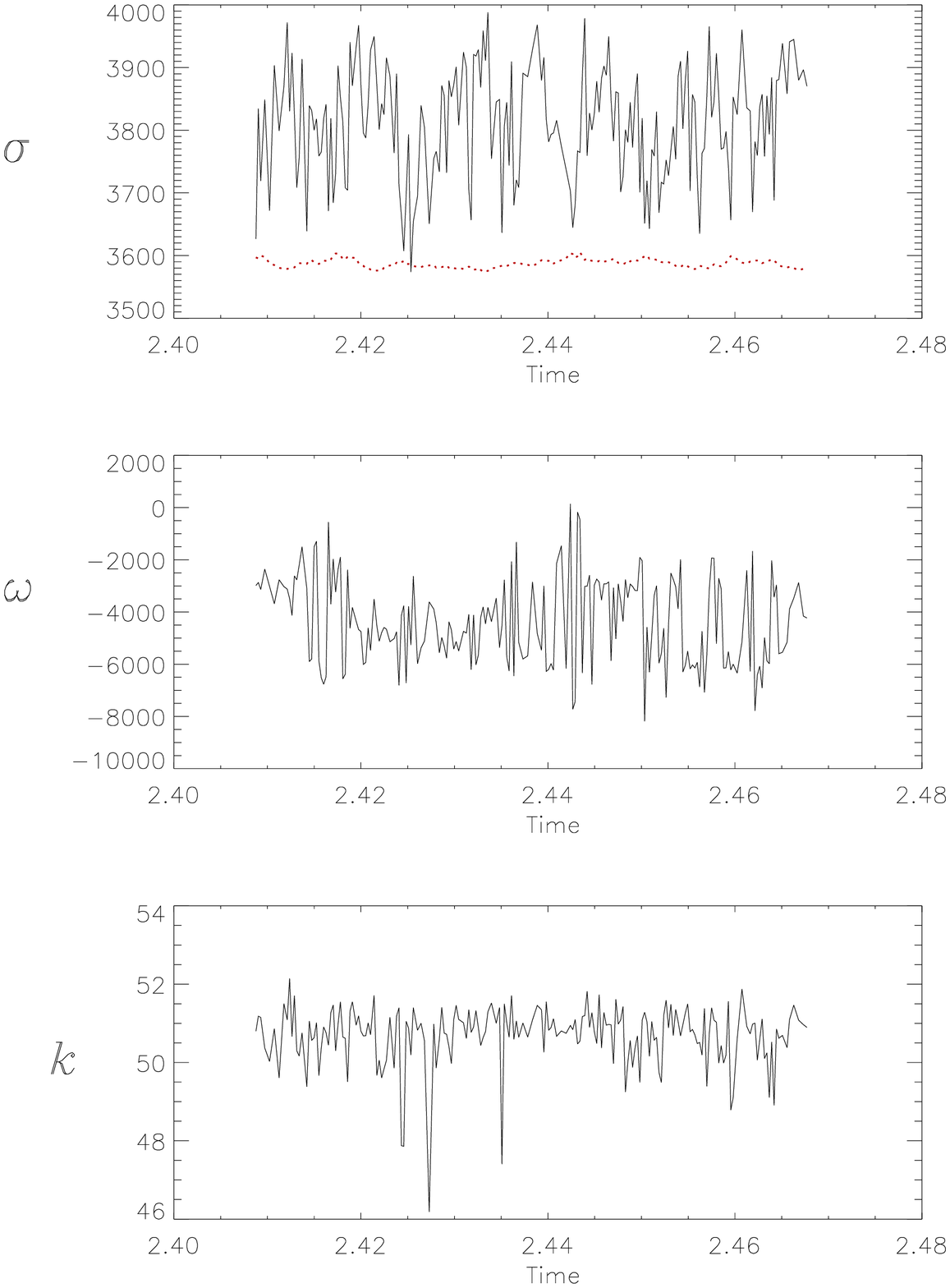}}
%\vspace{50pt}
\caption[Growth rate, frequency and wavenumber plots for runs with nonlinear zonal flow]{Time evolution of the growth rate, frequency and wavenumber plots for runs with only nonlinear zonal flow. 
The dotted line is the energy of the zonal flow.}
\label{fig:linzfs}
%\end{center}
\end{figure}

We begin with the case where \emph{only} the zonal flow, 
$\bar{U}$, is included in the linear theory. Hence in this subsection we set 
$U_0(y)=\bar{U}(y)$ and $G_0(y)=0$ in the linear equations 
(\ref{eq:psieqlin}-\ref{eq:thetaeqlin}).  
Figure \ref{fig:linzfs} shows how the growth rate, $\sigma$, frequency, $\omega$, and 
wavenumber, $k$, vary as the nonlinear system is evolved, for the runs for which plots were 
produced in figure \ref{fig:energies}.

Figure \ref{fig:XIIzf}, for run XII, can be compared with the plots of figure \ref{fig:XIIe}. 
The zonal energy of figure \ref{fig:XIIe} is shown as a dotted line to aid comparison.
As the zonal flow strength gradually decreases the quantities plotted in 
figure \ref{fig:XIIzf} remain fairly constant. However, there is a sudden increase 
in $\sigma$ and $k$ at $t\approx 0.247$, which is where $E_Z$ attains its minimum. This is 
expected as the growth of convection should occur when the zonal flow is weakest. Although 
the range of the growth rate is quite large, we notice that $\sigma$ is never less 
than $\approx 1500$. Therefore the zonal flow reduces the growth of the convection but does 
not completely cause it to cease. The zonal flow $E_Z$ increases strongly following
the burst of convection after $t\approx 0.247$, and  the growth rate 
begins to decrease again due to the disruption of the convection by the additional strength 
of the zonal flow.
%Figure 5

Unlike in the case for run XII, the growth rate in figure \ref{fig:VIIzf} remains relatively 
constant. The correlation with $E_Z$ in figure \ref{fig:VIIe} is also far less obvious, so it seems 
again that the zonal flow is not sufficiently affecting the growth of convection. There is 
excellent correlation however between the frequency, $\omega$, and the zonal flow strength. The 
frequency is smallest in magnitude when the zonal flow is weakest. Peaks in $k$ also coincide with 
locations of strong zonal flow although the range of the wavenumber is small. Run XV also displays 
bursting and again there is correlation between the quantities of figures \ref{fig:XVzf} 
and \ref{fig:XVe}. Once again the minimum growth rate is attained when the zonal energy is largest 
but the zonal flow is unable to reduce the growth rate to marginal or decaying modes. When comparing 
figures \ref{fig:XVIIzf} and \ref{fig:XVIIe} we immediately notice the lack of correlation between 
quantities that was present for the previous runs discussed and thus the departure from 
the $U_0=0$ case is minimal. This is to be expected since run XVII is not a bursting solution and 
is included here simply as an example of a non-bursting run.

We can conclude from this subsection that the zonal flows of the nonlinear theory certainly have 
a profound effect on the linear growth rates of convection. However, the zonal flow is unable to 
halt the growth of convection altogether as evidenced by the lack of negative growth rates in 
figure \ref{fig:linzfs}. Therefore another process, at least in part, must be responsible for 
the sufficient reduction in convective growth.

\subsection{Linear results with nonlinear mean temperature gradient}
\label{sec:nllintg}

\begin{figure}[p]
%\begin{center}
%\hspace{-25pt}
\subfigure[\ Run XII.]{\label{fig:XIItg}\includegraphics[scale=0.38]{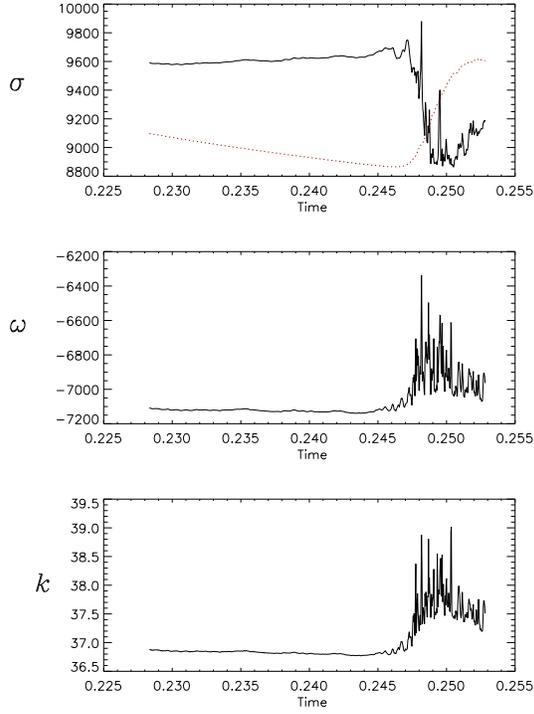}}
\hspace{30pt}
\subfigure[\ Run VII.]{\label{fig:VIItg}\includegraphics[scale=0.38]{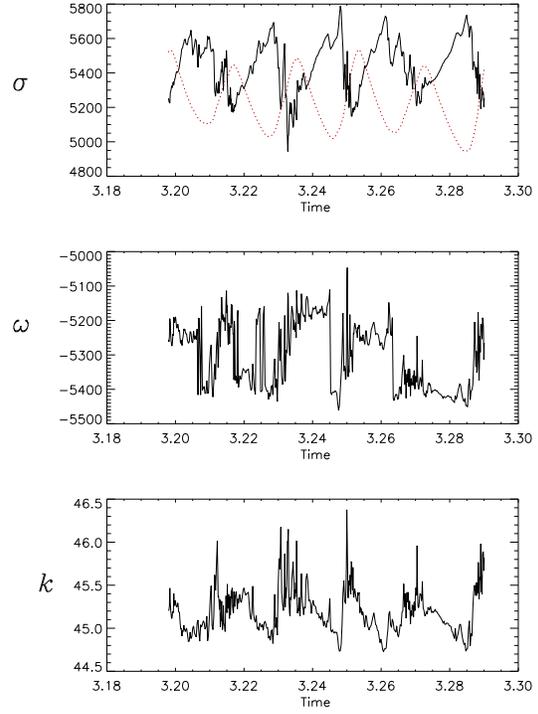}}
\subfigure[\ Run XV.]{\label{fig:XVtg}\includegraphics[scale=0.38]{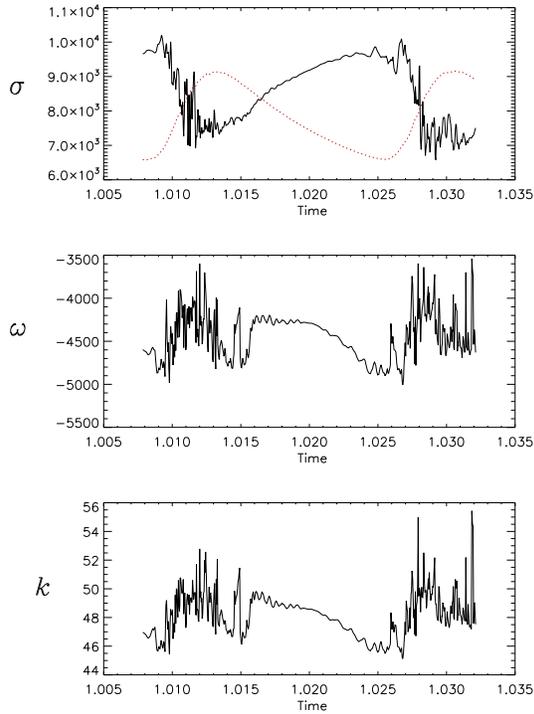}}
\hspace{30pt}
\subfigure[\ Run XVII.]{\label{fig:XVIItg}\includegraphics[scale=0.38]{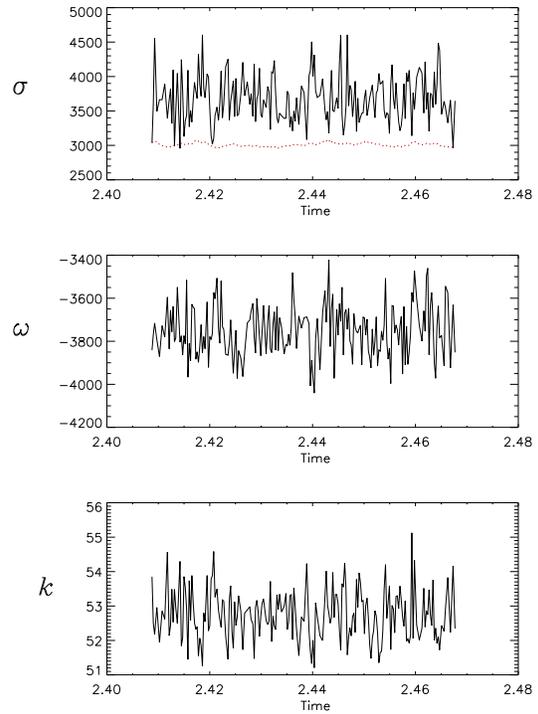}}
%\vspace{50pt}
\caption[Growth rate, frequency and wavenumber plots for runs with nonlinear mean temperature gradient]{Time evolution
of the growth rate, frequency and wavenumber plots for runs with only nonlinear mean temperature gradient.
The dotted line is the energy of the zonal flow, included since the mean temperature gradient is well correlated with
the zonal flow.}
\label{fig:lintgs}
%\end{center}
\end{figure}

We now consider the linear stability results in the absence of any zonal flow but with the mean 
temperature profile, $\bar{\theta}$, included. Thus, in this subsection we set $U_0=0$ and 
$G_0=\bar{\theta}$ in the linear equations (\ref{eq:psieqlin}-\ref{eq:thetaeqlin}). 
Figure \ref{fig:lintgs} contains plots displaying how $\sigma$, $\omega$ and $k$ vary 
as the nonlinear system is evolved when only the mean temperature gradient is included in 
the linear system. We show also the zonal flow energy, which varies smoothly and is 
well-correlated with the mean temperature gradient.

%Figure 6

All three of the quantities in figure \ref{fig:XIItg} remain near-constant to begin with since the 
extrema of the mean temperature gradient are also approximately constant for $t<0.247$ (compare with 
figure \ref{fig:XIIe}). The sudden increase in $\bar{\theta}^\prime_{\textrm{max}}$ at 
$t\approx0.247$ is accompanied by an abrupt reduction in the growth rate. This is to be expected 
since if the mean temperature gradient is able to partially (or indeed, fully) cancel out the 
static temperature gradient, the overall gradient will be less adverse. Thus the system will be 
less eager to convect, resulting in a lowering of the growth rate. However, even when the mean 
temperature gradient is strong the growth rate is only reduced by approximately $10\%$. In fact, 
this is a smaller reduction of the growth rate than was present in the previous subsection. 
Associated with the region of strong mean temperature gradient, there is a reduction in $|\omega|$ 
and the wavelengths of the modes.

The plots in figure \ref{fig:VIItg}, for run VII, show clear correlation with \ref{fig:VIIe}. The growth rate 
oscillates, though again does not reduce significantly. The correlation of the frequency and wavenumber is also 
clear with the same dependence as seen before. In figure \ref{fig:XVtg}, for run XV, we again see the same pattern 
of correlation by comparing with figure \ref{fig:XVe}. Peaks of $\bar{\theta}^\prime_{\textrm{max}}$ at $t\approx 1.012$ 
and $t\approx 1.029$ are associated with weak growth and short wavelengths whilst the intermediate period has 
increasing growth. The plots for run XVII, displayed in figure \ref{fig:XVIItg}, display only small fluctuations 
in $\sigma$, $\omega$ and $k$. This is to be expected since the values of the extrema of the mean temperature 
gradient are near-constant in this non-bursting solution (see figure \ref{fig:XVIIe}).

We have found that a strong mean temperature gradient can indeed reduce the growth rate of convection due to a 
reduction in the overall adverse temperature gradient present. However, the growth rate does not become marginal or 
negative even during times of strong mean temperature gradient. 

\subsection{Linear results with both nonlinear mean quantities}
\label{sec:nllinzftg}

We now finally consider the linear stability results with both mean quantities, $\bar{U}$ and $\bar{\theta}$, included 
in the basic state since we expect that both a zonal flow and a mean temperature gradient are necessary to produce 
the bursting phenomenon. Therefore in this subsection we set $U_0=\bar{U}$ and $G_0=\bar{\theta}$ in the linear 
equations (\ref{eq:psieqlin}-\ref{eq:thetaeqlin}).

\begin{figure}[p]
%\begin{center}
%\hspace{-25pt}
\subfigure[\ Run XII.]{\label{fig:XIIzftg}\includegraphics[scale=0.38]{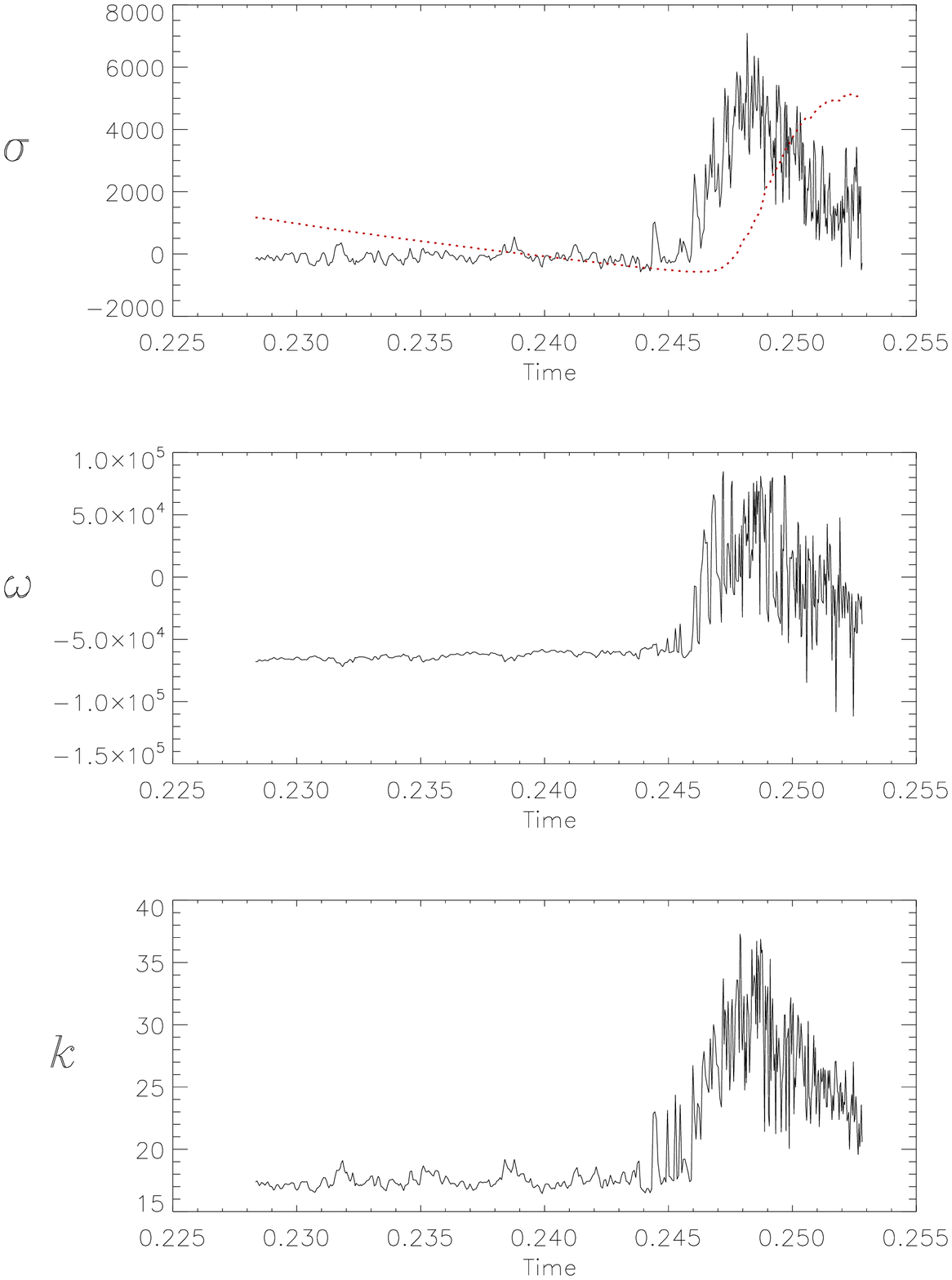}}
\hspace{30pt}
\subfigure[\ Run VII.]{\label{fig:VIIzftg}\includegraphics[scale=0.38]{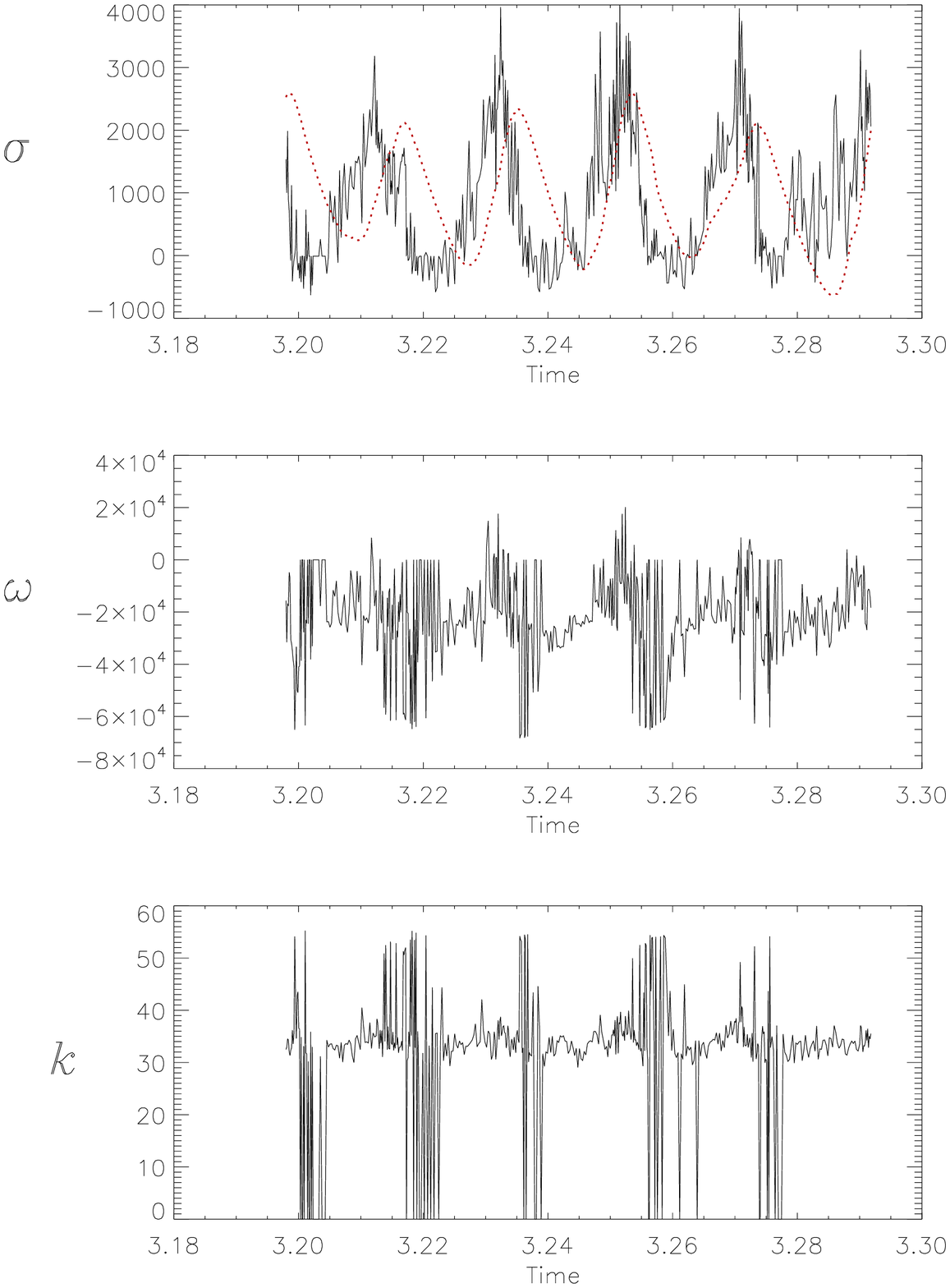}}
\subfigure[\ Run XV.]{\label{fig:XVzftg}\includegraphics[scale=0.38]{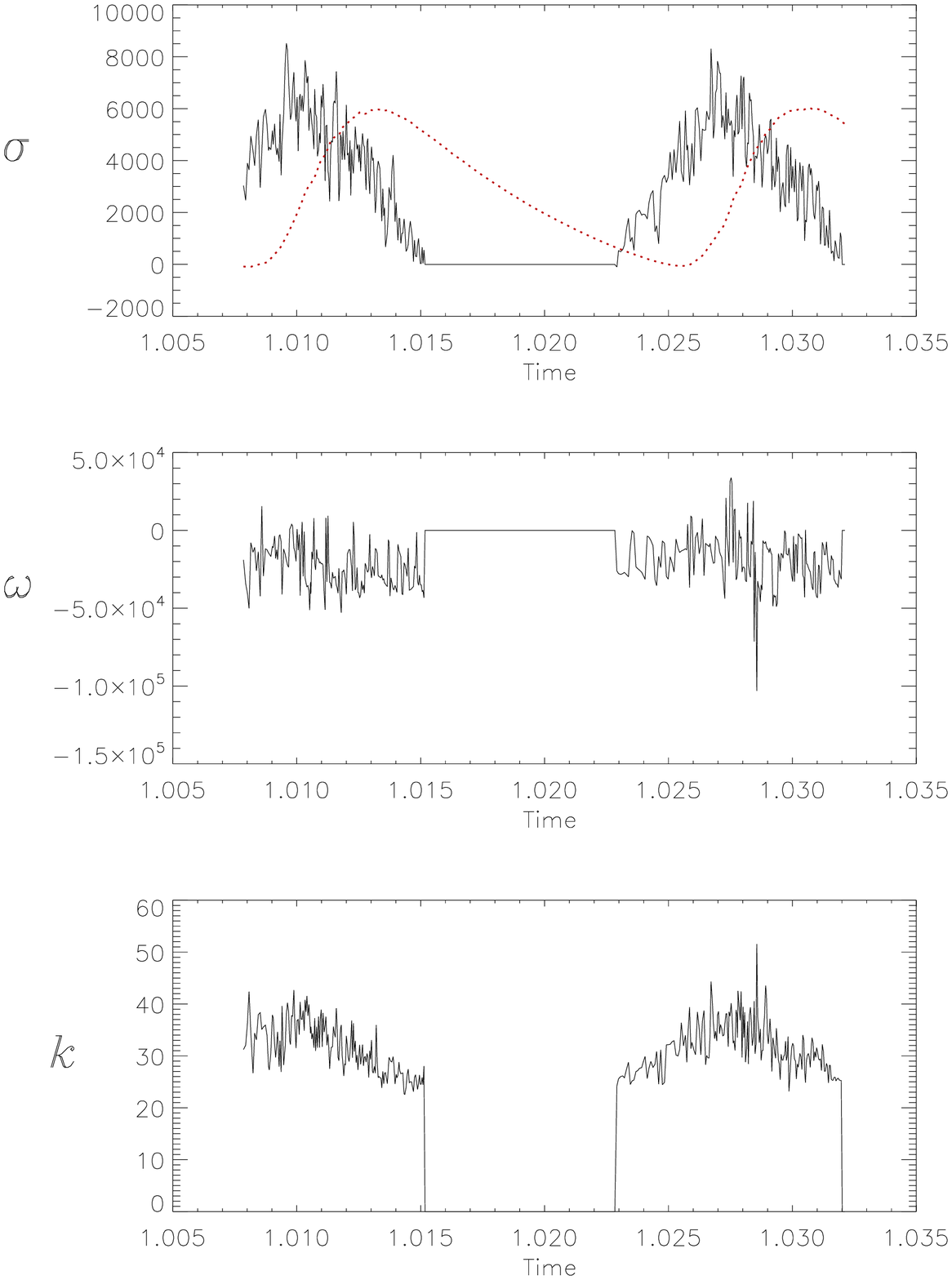}}
\hspace{30pt}
\subfigure[\ Run XVII.]{\label{fig:XVIIzftg}\includegraphics[scale=0.38]{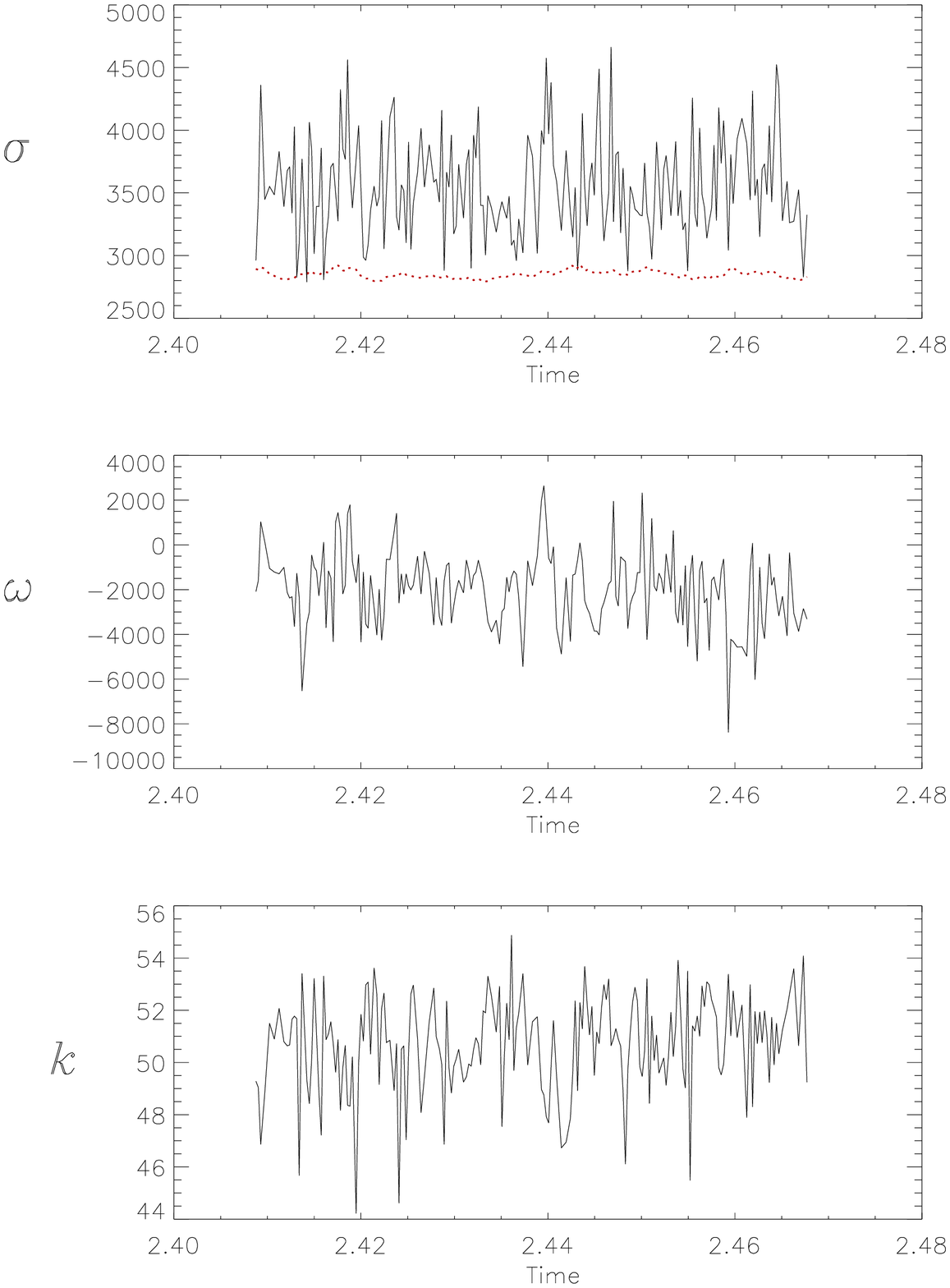}}
%\vspace{50pt}
\caption[Growth rate, frequency and wavenumber plots for runs with nonlinear mean temperature gradient]{Growth rate, frequency and wavenumber plots for runs with zonal flow and nonlinear mean temperature gradient.}
\label{fig:linzftgs}
%\end{center}
\end{figure}

The comparison of figure \ref{fig:XIIzftg}, for run XII, with figure \ref{fig:XIIe} shows that there is again 
correlation between the linear quantities and the nonlinear energies. In fact, the plots of figure 
\ref{fig:XIIzftg} are extremely similar to those of figure \ref{fig:XIIzf} where only a basic state zonal flow 
was included. Strong growth of the same order of magnitude remains possible at times when the zonal flow and 
mean temperature gradient are weak. However, the key difference between these sets of plots is that, for the 
case where both mean quantities are included, the growth rate is approximately zero when the mean quantities are 
large. This was not the case previously and therefore including both mean quantities has given the desired result 
which is the ceasing of the convection.

%Figure 7

The correlation of $\sigma$ in figure \ref{fig:VIIzftg}, for run VII, with the quantities plotted in figure 
\ref{fig:VIIe} is striking. As with figure \ref{fig:VIItg} there is strong growth located where the zonal flow 
and mean temperature gradient are weak. However, unlike figures \ref{fig:VIIzf} and \ref{fig:VIItg}, the growth 
rate becomes negative when it attains its minimum values. Hence when the mean quantities are large the convective 
modes of the linear theory decay. 
Figure \ref{fig:XVzftg}, for run XV, also appears to show that both mean quantities are necessary for bursting. 
There is an initial period of strong growth at $t\approx 1.010$ where we see from figure \ref{fig:XVe} that the 
mean quantities are weak. Followed by the strong growth there is a period where $\sigma\approx 0$ coinciding with the 
time between which $E_Z$ reduces from its maxima to its minima. After the zonal energy attains its minimum value, the 
zonal flow is weak enough to allow a second period of strong growth located at $t\approx 1.026$. Also of interest in 
both figures \ref{fig:VIIzftg} and \ref{fig:XVzftg} is that $k$ and $\omega$ tend to zero during periods of weak 
growth. The marginal modes, found when the mean quantities are strong, are therefore steady in these cases. The 
plots displayed in figure \ref{fig:XVIIzftg} are similar to those found for the non-bursting run XVII in the previous 
subsections. 
Once again all three quantities take (non-zero) near-constant values as expected, due to the weak mean quantities 
for run XVII.

We can conclude from this subsection that it appears that the necessary condition for bursts of convection is the 
existence of \emph{both} a zonal flow and a mean temperature gradient. We have observed marginal growth rates in all 
three runs that admit bursting. The Rayleigh number in all runs is several times critical. Thus, when the mean 
quantities are strong and of the correct form, they are able to reduce the system to near-onset behavior.

%%%%%%%%%%%%%%%%%%%%%%%%%%%%%%%%%%%%%%%%%%%%%%%%%%%%%%%%%%%%%%%%%%%%%%%%%%%%%%%%%%%%%%%%%%%%%%%%%%%%%%%%%%%%%%%%%%%%
\section{Conclusions}
\label{sec:con}

The results of our nonlinear annulus model produced good agreement with previous simulations \cite{jonabd03} 
and zonal flows were found to readily occur. Multiple jets and a periodic nature of convection appearing in bursts 
can be found under certain parameter regimes. However, bursting multiple jet solutions were not observed at any 
Prandtl number, extending the idea that multiple jets and bursts are likely to be mutually exclusive phenomena 
\cite{rotjon06} to cases with $Pr\ne 1$. Rigid top and bottom boundaries are preferable for multiple jets whereas 
bursts of convection certainly prefer stress-free boundaries. Zonal flows are also found to be weaker with rigid 
boundaries implemented. We also found fluctuations in the mean temperature gradient on a similar timescale to the 
bursts of convection which have not been addressed in the previous literature. We found reasonable agreement with the 
Rhines scaling theory \cite{rhi75} only when the zonal velocity was using in the scaling. It seems that the convective 
velocity is unable to predict the correct number of jets although further parameter regimes, including with larger 
values of $\beta$, should be tested confirm this result.

As an extension to the previous work, we performed runs with $Pr\ne 1$. In general, increasing the Prandtl number 
depletes the strength of the zonal flow. The bursts of convection appear to be a phenomenon most frequently observed at
$Pr=1$ agreeing with previous work \cite{chr02}. Although we found that bursts were possible at a range of Prandtl numbers, 
the onset of bursting appears to be delayed to larger Rayleigh numbers if the Prandtl number is not unity. 
This was most notably confirmed at $Pr=0.2$ by the lack of bursts at $Ra/Ra_c=2.75$ and the appearance of only very weak 
bursting at $Ra/Ra_c=5$. At even larger Rayleigh numbers the convection becomes highly chaotic. Therefore 
it appears that the bursting phenomenon may be restricted to an ever shrinking window of parameter space as the Prandtl number is reduced. 
At low enough Prandtl number the bursting regime may be omitted altogether restricting the phenomenon to a finite range of $Pr$.
This will have to be tested in future work.

Physically, the zonal flow certainly disrupts the convection as expected and as observed in section 
\ref{sec:nllinzf}. Similarly, the introduction of a strong mean temperature gradient can result in the reduction 
of the overall temperature gradient, $T^\prime = \Delta T/D + \bar{\theta}^\prime$. The adverse temperature gradient 
must exceed some value in order for convection to be beneficial. Also, the steeper the adverse temperature gradient 
the stronger the resulting convection will be. Hence a partial cancellation of the static temperature gradient, 
$\Delta T/D$, will also weaken the convection. We believe that the shearing of the zonal flow, coupled with the 
partial balancing of the adverse temperature gradient, is the requirement to halt convection. This is in contrast 
to previous work on the subject where it was believed that the zonal flow could sufficiently disrupt the convection 
to cause bursts. Both the zonal flow strength and the mean temperature gradient must also exceed some critical value 
in order for the convection to cease. In the case of the zonal flow the shearing must be great enough and in the case 
of the mean temperature gradient the static temperature gradient must be sufficiently balanced. When this occurs, the 
driving force of both of the mean quantities is removed. Consequently, there is a depletion in the strength of the 
zonal flow and the temperature gradient reverts to approximately that of the static case so that convection is once 
again beneficial and a burst occurs. This argument also offers an explanation as to why bursting is preferentially observed at 
Prandtl number of order unity. At high Prandtl number, the zonal flow is too weak
for bursting, and at low Prandtl number, although the zonal flow is strong,
the mean temperature is too close to its conduction state value. 

It is not currently known if the jets of the gas giants possess a periodic nature. The parameter regimes we have 
tested suggest it may be unlikely that the multiple jet structure of the Jovian atmosphere can coexist with bursts 
of convection. However, if the high latitude jets are driven by a different process to that of the strong equatorial 
jets \cite{hei05,hei07}, it may be that some but not all jets display an oscillation in the zonal flow strength. 
Further observations of the wind speeds of the jets of the gas giants over time is required. The Juno mission which 
launched in August 2011 will be placed in a polar orbit of Jupiter in order make further observations of the planet 
including their jet speeds \cite{mat07}.

%%%%%%%%%%%%%%%%%%%%%%%%%%%%%%%%%%%%%%%%%%%%%%%%%%%%%%%%%%%%%%%%%%%%%%%%%%%%%%%%%%%%%%%%%%%%%%%%%%%%%%%%%%%%%%%%%%%%

%\label{}

%\subsection{}

%\subsubsection{}

% If in two-column mode, this environment will change to single-column format so that long equations can be displayed. 


%\begin{widetext}

%$$\mbox{put long equation here}$$

%\end{widetext}

% Figures should be put into the text as floats. 

% Use the graphics or graphicx packages (distributed with LaTeX2e).


%



% Put the label that you will use with \ref{} command in the braces of the \label{} command.

%

% \begin{figure}



% \end{figure}







%

% \begin{table}



% \end{tabular}

% \end{table}


\begin{acknowledgments}

RJT is grateful to STFC for a PhD studentship.

% Put your acknowledgments here.

\end{acknowledgments}


%\bibliographystyle{unsrtnat}
%\bibliographystyle{plainnat}
%\bibliography{paper4}

\end{document}